# Spontaneous Emulsification: Elucidation of the Local Processes

Monicka Kullappan[1,3], Wes Patel[1] and Manoj K. Chaudhury[1,2]

[1]Department of Chemical and Biomolecular Engineering

[2]Department of Materials Science and Engineering, Lehigh University

Bethlehem, Pennsylvania 18015, United States

[3]Current Address: GE Research, Niskayuna, New York 12309. USA

**Abstract**. Micro and/or Nano sized emulsions are formed when an organic liquid gently comes in contact with water in the presence of a surfactant, where no external agitation is required. Many years of research made it clear that the driving force for spontaneous emulsification arises from the differences of the chemical potentials of various components in the organic and aqueous phases, which triggers diffusion coupled hydrodynamic fluctuation. While extraordinary theoretical developments have taken place that attempted to describe these processes within the scopes of equilibrium and non-equilibrium thermodynamics, the local processes underlying the spontaneous emulsification, however, still remain elusive. In this research, we investigate the local processes that involve the transfer of surfactant as well as water from one phase to another (i.e. water to oil), which results in the formation of water-in-oil emulsion in the organic phase and, subsequently. Thes emulsions invert into oil-in-water emulsion, rather abruptly, as they cross the phase boundary. Studies based on UV spectroscopy and molecular dynamics indicate that these processes may involve explosive events and subsequent assembly of the fragments to other organized structures which are reminiscent of cusp catastrophe proposed earlier by Dickinson. These processes lead to either to a strong or a weak fluctuation of the component concentrations below the interface that also becomes evident in the fast (athermal) diffusion of the emulsion droplets from the interfacial region farther into the bulk water. These events can be arrested suitably with polymeric additives.

## Introduction

Microscopic or submicroscopic emulsions[1] are dispersions of droplets of one liquid into another that can be formed via agitation or even without it. Spontaneous emulsification, in which no agitation is required, was first discovered by Gad, followed by a series of papers by Quincke, Brucke and others, which were documented by McBain and Woo[1]. While our understanding of agitation induced emulsification has advanced considerably since the original publication of Hinze [2], the processes that lead to emulsification without agitation[1, 3-21] continue to inspire new imaginations even after many decades of research in this field. The fundamental question that the researchers had to grapple with is how is it possible to create new surface areas associated with emulsification, which, of course, would be trivial to understand had the oil-water interfacial tension been negative [7,8]. While this scenario cannot be ruled out in a few specific cases, spontaneous emulsification, nonetheless, occurs predominantly with positive interfacial tension[7]. A conventional wisdom in this field is that the increase of the interfacial free energy due to the formation of multiple droplets is somewhat compensated by an increase of the entropy of dispersion. In other words, emulsions can form[7] if the net change of free energy is negative. Spontaneous emulsification is generally accompanied with various spectacular dynamic effects such as the kicking and oscillation of pendant drops, twitching and sudden bursts of an interface, and rapid penetration of droplets in a liquid against gravity. These phenomena are now well-known and well documented in a plethora of publications[3,4,14,15].

It has been realized over time that the hydrodynamic instabilities can develop at the interface that could originate from Marangoni instability due to local variation of interfacial tension at the interface. This may occur via spontaneous fluctuation of concentration , which, according to McBain et al[1,3], was first conceived by Quincke. The chemical potential gradients across the interface lead to mass transfer and a concomitant hydrodynamic instability[3], whereas an in-plane gradient of concentration can lead to Marangoni force induced hydrodynamic flow. Due to a resurgent interest in this field of research, new insightful papers have continued to be published even recently[17, 22-28], whereas detailed accounts of the earlier research have been reviewed[16,18]. As this particular article is not meant to be a review, here we summarize only the salient features of the related literatures.

One important breakthrough that has occurred in understanding spontaneous emulsification is understanding the role of the so-called Interfacial Turbulence, which is the consequence of mass transfer coupled with solutary Marangoni flow near and at the interface. McBain et al[1,3] presented some of the earliest observations. With experiments performed with drops of xylene placed on aqueous surfactant solution, the authors observed that the droplets either explode violently or numerous small droplets are emitted from the larger droplet. The authors proposed an important concept in that the spontaneous emulsification results due to the self-generated eddies, which tear off large droplets of the oil. This picture is similar to that presented later by Hinze[2], but in situations where the eddies are generated by agitation induced turbulence.

McBain et al[3] also proposed that the driving force for spontaneous emulsification in system with positive interfacial tension arises from the energy of adsorption, as well as the solubilization of hydrocarbon in the aqueous detergent solution. A rigorous theoretical treatment of interfacial turbulence was provided later by Sterning and Scriven[15], who performed a stability analysis of an interface across which mass transfer occurs. The base state of the problem defines the balance of the viscous and the Marangoni stresses. The bifurcation analysis can then be performed about the base state by allowing a fluctuation of surfactant concentration at the interface. At the scaling level, the interfacial flux can be described in terms of such well-known dimensionless parameters as Marangoni and Schmidt numbers. Golovin[29] stated two special scenarios. The first one is chaotic, in which disturbances are local and where the fluctuations die out fast following such sporadic activities as pulsation, bursts and twinkling of the interface. Another scenario results with a slower dissipation of energy, in which periodic roll cell like flow patterns are formed below and above the interface[14,]. Sherwood and Wei[4] as well as Gooh Pattader et al [30,31] have provided further exposition of the roles of such interfacial instabilities in various mass transfer processes, which have immense implication in liquid-liquid extraction technologies.

While interfacial convection results from Marangoni induced hydrodynamic instability, one may ask if such a convection in itself, is responsible for emulsification of one liquid phases into another. In order to explore this point, Miller et al[10] added surfactant to such an extent that the Marangoni number tended to zero. However, even in such conditions, emulsions were formed spontaneously. This led Miller et al[10-12] to suggest that that spontaneous emulsification is primarily a result of chemical rather than a mechanical instability.

Following a line of argument proposed earlier by Kirkaldy[32] in the context of crystallization in a three-component system, Miller et al[10-12] suggested that the concentration of a diffusively transferred species across an interface first reaches a supersaturation, from where stable phases are formed via spinodal process. This idea of the formation of an organized structure from a supersaturated solution was proposed originally by Ostwald[33] in the context of crystallization. See also the work of Ostrovsky and Good[9] , and the references cited therein, which illustrates the application of this idea to emulsion formation. In order to exemplify the scope of chemical instability and from there the phase transformation, Friberg[34] mapped the evolution of the concentration of the species on a ternary phase diagram, which allowed visualization of which of the paths lead to single- or two-phase regions. Recently, Neibloom et al[28] extended similar ideas to spontaneous emulsification involving chemical reaction and elucidated the roles of diffusion and stranding underlying such processes.

The analysis of Sternling and Scriven[15] was restricted to the instability of planar interfaces, while, in reality, the interface can deform[35,36]. The concomitant refraction of hydrodynamic streamlines can create local vortices and hydrodynamic instability that may ensue even in the absence of a surface tension gradient. The deformation of interface poses difficulties in tracking the evolution of a system as various paths are plausible. Kirkaldy[36] proposed that this problem can be resolved using the principle of non-equilibrium thermodynamics, in which the preferable paths are those that minimize the rate of entropy production.

There, of course, remains the possibility that all the events discussed as above may participate concurrently in smaller or greater degrees in an emulsification process. Furthermore, if an oil containing a surfactant is brought into contact with an aqueous phase, water may diffuse in the oil triggering an aggregation of a surfactant and then its precipitation out of the solution, either as crystal or amorphous aggregate. When these precipitates come into contact with the oil-water interface, it would deliver surfactants to the interface rendering it to be locally supersaturated. During the relaxation of the interface from the local supersaturated heterogeneity, it could generate interfacial shock wave, local fluctuation of interfacial area, and folding/budding transition[37,38]. These could also trigger pulsation, twinkling and bursts, which are ubiquitously observed interfacial events.

The brief discussion as presented above highlights the richness and complexity underlying spontaneous emulsification, the understanding of which requires adequate blends of interfacial science, transport phenomena, phase evolution kinetics within the domain of non-equilibrium thermodynamics. It may be naïve to presume that a general understanding of spontaneous emulsification is imminently forthcoming, although certain semi-empirical approaches, such as those based on the calculation of HLB[1] (hydrophobic lipophilic balance) and/or the HLD[39] (hydrophobic lipophilic difference) numbers for a combination of surfactant, oil, water provide useful insights into such mechanisms. Additionally, the sudden transformation of an oil/water to water/oil emulsions or vice versa, which are recognized to be parts of the overall emulsification hierarchy can be captured using the idea of cusp catastrope[40-42]. However, this method, understandably, does not provide the much-needed mechanistic insights into the local processes as it does not claim to be anything more than a phenomenological model. We are motivated by the hypothesis that understanding the local processes, including those involving singular transformations of one phase to another, is important to develop a comprehensive understanding of the emulsification processes.

**Background, Motivation and Objective**

During the last five years, we have been conducting a research, the purpose of which is to elucidate how emulsions are formed in the ship bilges that occur, sometimes, under vigorous agitation, and, other times without it. This is a vexing problem for the armed forces, as all the naval ships collectively produce massive amount of oil contaminated water that is difficult and/or expensive to purify before the purified water can be discharged back to the ocean[43].

The complexity of bilgewater, which is a wastewater containing oils, fuels, salts, lubricants, deck and bilge cleaners, is very rich. An important question here is: how are the oil-in-water emulsions formed in bilgewater, where the concentration of the surfactant is typically close to, but slightly lower than, the critical micellar concentration. This particular fact that micelles and, therefore, emulsions can form at a concentration below c.m.c is, in itself, not as striking as it might have appeared before the publication of a seminal work by Besseling and M. A. Cohen Stuart[44]. Using self-consistent field theory these authors demonstrated that the above-mentioned structures can form via nucleation. It is only that their formation becomes unchallenged at c.m.c when the energy barriers disappear, which is a general knowledge in emulsion and polymer sciences. The daunting challenge with the bilgewater stems from its highly complex composition. Nevertheless, it is possible to identify the minimum set of parameters with which the complexity of the problem can be preserved in sufficient level with which to develop a rational understanding of the emulsification process in bilge water (see reference 26 and the references cited therein)

In our previous publication, we primarily focused on emulsification under agitated conditions that lead to hydrodynamic turbulence[26]. Here, we pay our attention to what happens without any external agitation. The system that we use here is same as that published previously: specifically, it involves an equi-weight mixture of an ionic (dodecanoic acid) and a non-ionic (Brij 35, Polyoxyethylene (23) Lauryl Ether) surfactant, dissolved in the 0.6 M aqueous salt solution over which various purified organic liquids were gently deposited. [Note: In what follows, we use the term oil for the organic phase intermittently for convenience although the term "organic phase" is more general and used most often]

What would be immensely useful in interrogating the dynamic and the self-organizing behaviors of the surfactants is to exploit a property of the surfactant itself that would be *reporting* the states that they evolve through. In this regard, we mention here a recent elegant investigation[46], where the authors used a solvatochromic surfactant, the fluorescent intensity of which depends on the polarity of the surrounding environment. This particular characteristic allowed a dynamic visualization of the surfactant dynamics at and near the oil-water interfaces.

In the similar vein as these previous studies, here we used ultraviolet (UV) spectroscopy that allowed us to discern whether the surfactants are in free or in aggregated state in both the organic and aqueous phases. This was possible as the UV absorbing carboxylic acid groups exhibit Bathochromic (red) and Hypsochromic (blue) shifts depending upon whether they form aggregates via H-bonding interaction with each other and/or with the solvent itself [47,48]. Information of such interactions in both the organic and aqueous phases has been vital in our understanding of the formation of various local processes. For example, if the relevant chromophore of a surfactant exhibits a red shift in a non-polar environment, we may say that this is possibly due to intermolecular H-bonds, thus inferring the presence of dimeric, trimeric or polymeric states[47,48] that they are in. Higher order shifts would also be possible in a non-polar solvent if the latter contains a small amount of water thus promoting the formation of such a self-assembled structure as an inverted water-in-oil emulsion. When the surfactant is present in aqueous phase, other types of chromic states could also be observed due to the polybasic character of the carboxylic acid[49,50], which could be interrogated via pH dependent Bathochromic titration (see below). While the UV spectroscopy allowed us to interrogate various types of self-assembled structures that are formed, these studies, when it is carried out in parallel with a dissipative particle dynamics[51] (DPD) and/or combined with a molecular dynamics (MD) simulation, we gained even a deeper insight into these structures and their stability.

The basic experiment consisted of an organic phase (alkane, aromatics and diesel) that is deposited on the surface of an aqueous phase. Salt (sodium chloride, 0.6 M) was dissolved in water to produce a salinity comparable to that of seawater, following which an anionic surfactant (dodecanoic acid), a nonionic surfactant (Brij-35) or their mixture was dissolved in it at a total concentration of 100 ppm. These compositions closely parallel to those recommended for the simulation of bilge water.[27]

What we learned, at one level, is that the surfactants as well as molecular water diffuse into the organic phase from the aqueous phase below it. The surfactants then self-assemble in the organic phase to form inverted (water-in-oil) emulsion that then fuses with the interface thus forming an oil-in-water emulsion. These experimental results coupled with those obtained from the DPD based Molecular Dynamics using the Large-scale Atomic/Molecular Massively Parallel Simulation (LAMMPS) software, provided a more complete picture into the emulsification processes. Another insight gained from these studies is that the micelles and vesicles that are formed in the aqueous phase fuse intermittently with the interface explosively, following which various fragments are produced. This fragments then re-combine and lead to the formation of inverted water-in-oil emulsion in the organic phase. Subsequently, the inverted emulsions reach the interface from the organic phase and disintegrate again explosively following which the fragments cluster together and form oil-water emulsions.

**Roadmap.** We present the hypotheses, brief experimental protocols, and the associated simulations performed in the Results and Discussions section under several interconnected sub-sections. These are so arranged that the reader would be able to grasp the main message of each sub-section without the need to know the details of the experimental and the simulation procedures *a priori*. Details of the experimental and the simulation methods, nevertheless, are provided in the Supporting Materials Section, in step by step instructions. The subsections are arranged and written in such a way that when one theme reaches a closure, it prompts to the next one. This is the best way we could present this research, which has several interconnected stages of hypotheses and deductions, one leading to another. After consolidating the main points of this research, several references are cited, although omitting, regretfully, many important references.

## Results and Discussion

### Bathochromic and Hypsochromic Shifts: The Polybasic Character of Carboxylic Acid

There exists a plethora of literature[47-50, 52-54] that elucidate the various ways alkyl carboxylic acids interact with each other and with water (Figure 1). For example, they can form H-bonded cyclic dimer, not only in water but also in a non-polar liquid[53] .

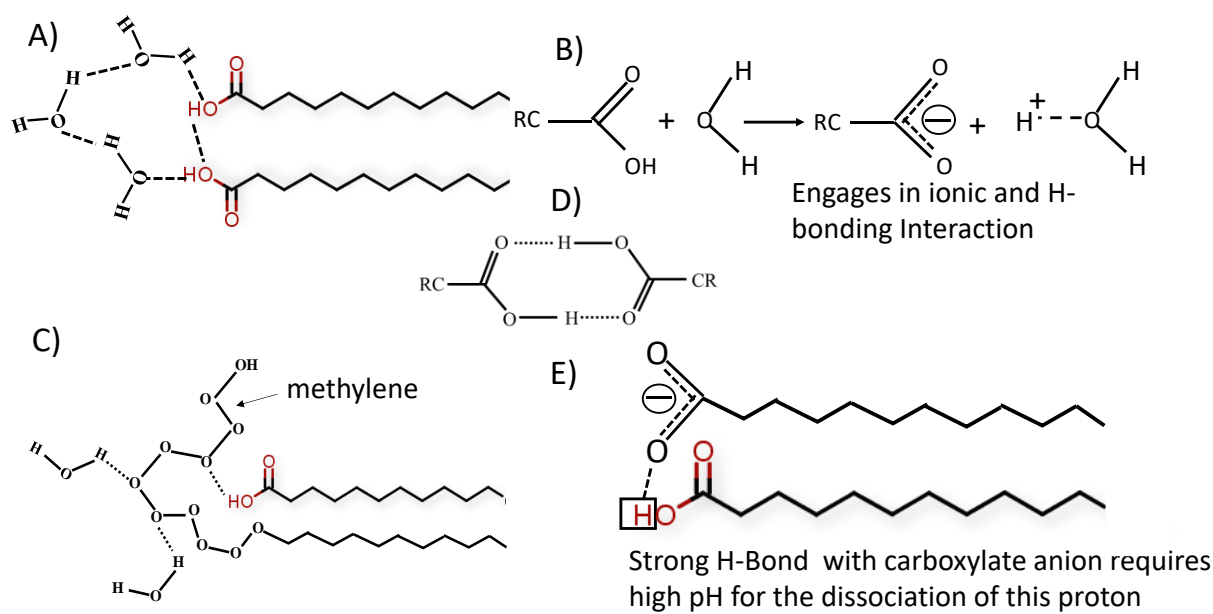

**Figure 1.** Various types of H-bonding interactions exhibited by carboxylic acid. A) Two carboxylic acids can interact with each other directly via H-bonding interaction or B) through H-bonding interaction with water [47]. C) Carboxylic acid interacts with the ether group of a non-ionic surfactant via H-bonds. D) Two carboxylic acid groups forming dimer E) A deprotonated carboxylic acid

acquires a net negative charge, which can interact with a neighboring protonated group via H-bonding interaction.

While two fatty acids can self-associate via hydrophobic attraction[52.54] of the alkyl groups in water, the association is strengthened due to additional H-bonding interaction with each other, either directly or indirectly via H-boding interaction with water. The carboxylic acid groups can also de-protonate thus forming carboxylate ion, which can also form various types of H-bonds with water or with another proton donating group. The ether group of a non-ionic surfactant can also engage in an H-bonding interaction with the protons of water or the carboxylic acid.

Ruderman et al[47] published an important paper, where they demonstrated via UV absorption spectroscopy and molecular dynamics simulation that acetic acids form dimer, trimer and higher order aggregates by forming H-bonds with each other and/or with water. The signature of the extended H-bond formation was evident in UV spectroscopy that exhibited progressive Bathochromic (red) shift with the concurrent increase of the aggregate size that occur when the concentration of the acetic acid is increased. This knowledge has been very crucial in interpreting the Bathochromic or Hypsochromic behaviors of surfactants, notably when carboxylic acid is a functional group of the surfactant.

**Surfactants in Aqueous Phase**. Similar to those reported by Ruderman et al[47], we too observed that the UV absorption peak of the carboxylic acid exhibits a progressive red shift (Figure 2 A) as the concentration of dodecanoic acid is increased in an aqueous phase from 2 ppm to 100 ppm. When these studies are carried out by changing the pH of the aqueous solution, the primary absorption peak (210 nm) undergoes a pronounced red shift starting from about a pH of 6 (Figure 2B). After a more basic pH (i.e. in the range of 9 to 12) is achieved, no further shift occurs.

Presumably, these red shifts occur due to the H-bonding interaction between the unprotonated and deprotonated carboxylic acids in an extended network as suggested by Katchalsky et al[50].

The fact that the density of the carboxylic group plays a role in the Bathochromic titration is evident in that the shifts are much smaller in magnitude when the concentration of the surfactant is decreased to about 20 ppm (Figure 2). From the midpoint of such a titration, the pKa of the carboxylic acid is found to be about 7.5, which is larger than that (5.2) expected of this organic acid in its native state. Similar behavior, i.e., the shift of the pKa to greater than 5.2 (Figure 2C), has been reported in the past[49] with the so called "contact angle titration experiments" in which water drops of various pH were deposited on a carboxylic acid functional surface of polyethylene.

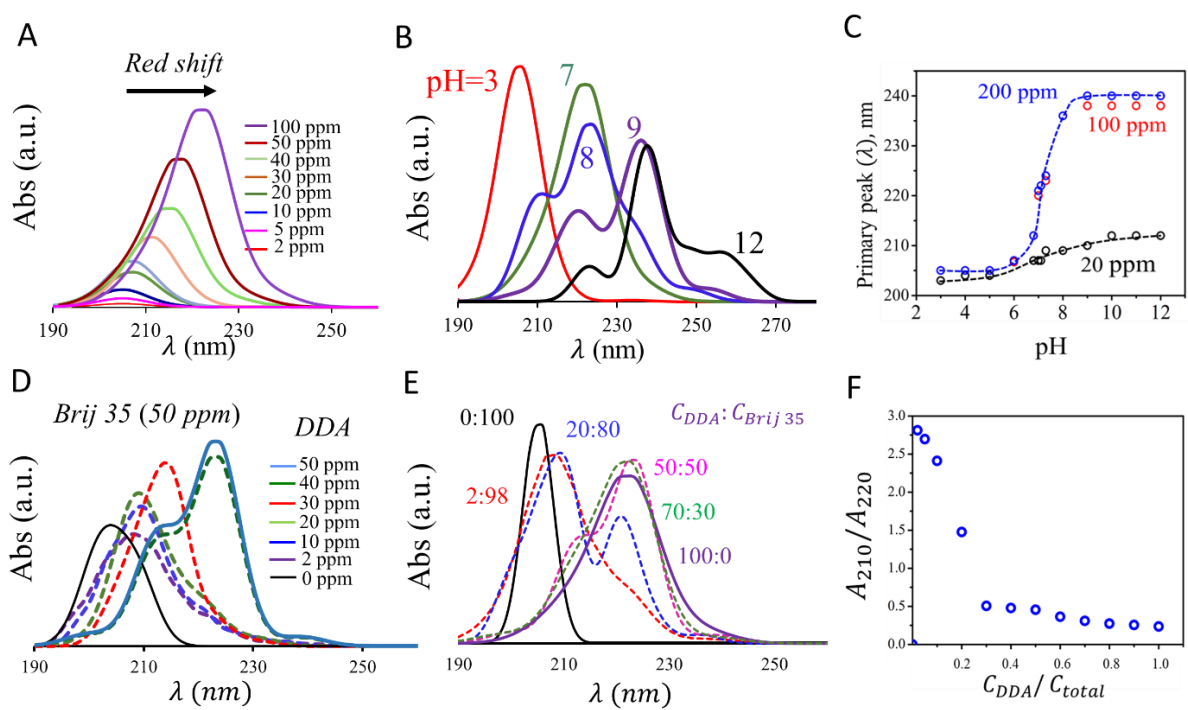

**Figure 2.** A) UV absorption spectrum of dodecanoic acid (*DDA*) exhibits progressive Bathochromic (red) shift as the concentration of the surfactant is increased. B) UV absorption spectra of dodecanoic acid exhibiting Bathochromic shift as pH of the aqueous phase increases.

C) A titration plot is obtained by plotting the primary absorption peak as a function of pH. D) UV absorption spectra of a mixture of dodecanoic acid and Brij-35 (Polyoxyethylene (23) Lauryl Ether), where the concentration of a non-ionic surfactant (Brij 35) is kept at a constant value (50 ppm), whereas the concentration of DDA is increased from 0 to 50 ppm. E) The ratio of Brij 35 to DDA is altered by keeping the total concentration to 100 ppm.  F) The ratio of peak heights at $\lambda$ = 210 nm and $\lambda$ = 220 nm decrease till the concentration of DDA reaches to 30 ppm, beyond which this ratio changes only slowly.

After performing the experiments without salt and with 0.1 M salt in water, the authors of that study[49] found no discernable dependency of the pKa on salt concentration.

Although we did not perform the Bathochromic titration experiments by varying salt concentration, we cite two previous observations[54] that support that the basic deprotonation behavior of carboxylic acid remains preserved even at high salt concentration. One of those observations pertain to the interaction of oil droplets (containing dodecanoic acid) with an amine functional substrate as studied using contact angle measurements, These studies performed in 0.6M salt solution as a function of function of pH showed that  the oil droplet barely adhered to the aminated substrate at a pH that is less than 6 or more than 10, whereas the adhesion was most tenacious at a pH near pH=8, where the de-protonated dodecanoic acid could interact with the protonated amine.   In another related experiment, it was found that the efficiency of the extraction of DDA containing Oil-in-water emulsion with amine functionalized particles was maximum at a pH close to 8.   We emphasize that those previous studies are consistent with the current Bathochromic titration results in that the carboxylic acid groups do deprotonate in 0.6 M salt solution, and it does so at a pH in the vicinity of  7.5.

It is known from the previous studies[47,50] that the polybasic acids ionize at a pH higher than that of its native pKa, as the deprotonation of an acid group becomes progressively inhibited by some of the ionized groups via electrostatic stabilization[49]. We note the carboxylic acid at very low pH (i.e., pH=3), exhibits a hypsochromic shift relative to that at pH 7 as the probability of the formation of electrostatic stabilization is reduced substantially at pH<< pKa. The main claim that we would like to make is that the Bathochromic shift of the carboxylic acid chromophore at a pH~7 and its polybasic character suggest that the dodecanoic acids are in close proximity to each other thus forming H-bonds, as would be expected if the surfactant exists in a self-assembled structure e.g. a micelle or a vesicle.

**Cooperative interaction between DDA and Brij 35.** In order to further interrogate the physical chemistry of the extended H-Bond formation, we investigated what happens when a nonionic surfatant is mixed with an anionic surfactant. The data summarized in Figure 2 (D and E) show that the dodecanoic acid exhibits a red shift in water when mixed with Brij 35 even when the latter is present at a rather low concentration (2 ppm), which is evident in two related but separate experiments. In the experiment, Bij-35 pre-existed in the aqueous phase at a concentration of 50 ppm, DDA was then added starting with a concentration of 2 ppm and ending to about 50 ppm. While DDA by itelf does not display an absorption peak > 205 nm at a contration of 2 ppm, it shows absorption peaks at higher wavelengths in response to the increase of the concentration of the anionic surfactant (Brij-35).

In the second experiment, the consecntration of both the surfactants in the aqueous phase were varied from 0:100 to 100:0 ratio. In Figure 2 (F)**,** we plot the ratio of the intensities of the peaks corresponding to 210 nm and 220 nm as function of the concentration of DDA in such a mixture. Here, a precipitous drop of this ratio occurs as the concentration of DDA is increased from 2 ppm to about 30 ppm, beyond which the ratio changes only slowly. These results are consistent with the scenario presented above that DDA engages in H-bonding interaction with Brij-35, and that a critical percolation threshold exists in such type of co-operative H-bond formation. More results related these observations can be found in the supplementary materials section (Figure S3 and S4),

**Surfactants in organic phase**. We now explore what happens when the surfactants are dispersed in an organic phase. The UV-spectroscopic studies were conducted with pure DDA, Brij-35 and their mixture in dodecane. Figure 3 shows that the peak corresponding to 100 ppm of DDA, which is centered around 210 nm, is rather broad. This is consistent with the existing literature[53] that carboxylic acid self-associates even in an apolar solvent. One would expect that the probability of dimerization would be rather low if DDA is present at a low concentration (~0.1 ppm). This result is indeed borne out, as Figure 3 shows that the UV peak of the carboxylic acid becomes narrower, and it exhibits a hypsochromic shift, relative to that at 100 ppm concentration. We expected that a similar hypsochromic shift should occur in the aqueous phase as well, provided that its concentration is rather low and if the study is conducted at a low pH (i.e. ~3), where both the H-bonding with the protonated and de-protonated carboxylic acid would be diminished, which is, indeed, borne out in the experimental observation. Figure 3A shows that the UV peak of DDA dispersed in the aqueous phase at pH 3 is rather close to that of DDA in dodecane when its concentration is as low as 0.1 ppm. The mixture of DDA and Brij-35 (50 ppm:50 ppm) in dodecane shows an absorption peak at 210 nm, which is similar to that of 100 ppm DDA in dodecane (compare Figures 3A and 3C). The peak at 205 nm for the Brij-35 (Figure 3B) virtually disappears in the mixture as it merges with that of the dodecanoic acid.

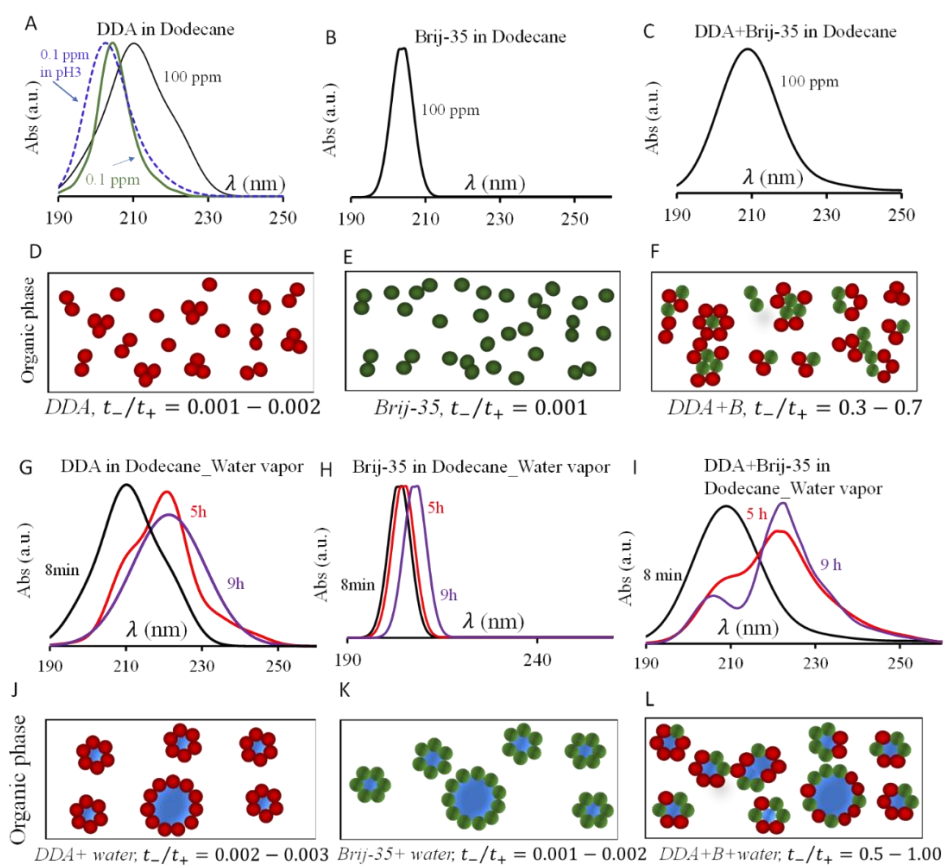

**Figure 3.** UV spectra of (A) Dodecanoic acid, (B) Brij-35 and (C) their mixture (50:50) in dodecane in the absence of water. In Figure A, the spectrum obtained for 100 ppm of dodecanoic acid is compared with 0.1 ppm dodecanoic acid in dodecane and that in aqueous phase at pH=3. DPD-MD simulations (D-F) show that dodecanoic acid molecules aggregate at least up to a trimer on its own, while Brij-35 formed dimers (E) under similar conditions. However, together with Brij-35, larger cluster (F) are formed. Panels (G-I); water vapor is infused into the samples. DPD-MD simulations show that inverted emulsion can form with each of the surfactants Panels (J-L), the stability of any of the aggregates can be discerned from the ratios of $t_-$ and $t_+$ (see text). Color coding are as follows, Red: DDA, Green: Brij-35 and Blue: water.

Let us contrast these observations with what we learned from the hydrodynamically coupled dissipative particle dynamics (DPD) coupled with molecular dynamics (MD) simulations. In order to find out the clustering abilities of the surfactants, the DPD-MD simulations were carried out . with either 100 ppm of DDA, 100 ppm of Brij-35 or a 50 ppm:50 ppm mixture of the two surfactants in dodecane. In a different set, these simulations were repeated in the presence of a small amount (100 ppm) of molecular water. Further details are provided in the method section, The results of these simulation, as summarized in Figure 3 (D - F), show that DDA can form weak dimers or trimers in dodecane. While Brij-35 forms a few dimers in dodecane, larger complexes are formed when DDA and Brij-35 are present together, in which the DDA clusters around Brij-35. These DPD-MD simulations also allow an estimate of the time ($t_+$) of formation of an aggregate and the time $(t_-)$ to its dissolution. The ratio $(t_-/t_+)$ is thus a measure of the stability of the aggregates, which is very low for both pure DDA (0.002) and pure Brij-35 (0.001) but is rather high (0.3 to 0.7) for the mixture of the anionic and nonionic surfactants. Overall, we see the picture developed from the UV spectroscopy contrasts positively with that obtained from the DPD-MD simulations in that an aggregated structure formed from a mixture of the ionic and non-ionic surfactants is more stable than those of the pure surfactants.

**Role of Moisture**. After gaining the needed insights into the interactions of the surfactants in pure dodecane, we next turn our attention to the role of a vapor of water infusing into the oil phase and the ensuing stability of the aggregates thus formed. To achieve this objective experimentally, water vapor saturated air was made to bubble through dodecane with surfactants pre-dissolved in it. As time progresses, it is observed that the primary peaks of the surfactants shift towards the higher wavelengths (Figure 3 (G - I)). The peak for DDA corresponding to 210 nm decreases, whilst that at 220 nm increases. These bathochromic shifts mark the role of water in the formation of H-bonded structures, which would be possible if the surfactants are in close proximity to each other, as would be the case if they form oil-in-water emulsions. DPD-MD simulations Figure 3 (J to L) clearly show the possibility of the formation of water-in-oil emulsions with pure and mixed surfactants. However, these inverted emulsions exhibit different degrees of stabilities, as indicated by the ratio $t_-/t_+$ . Here, too, as is the case with dry dodecane, the mixture of the anionic and nonionic surfactants is more stable than those formed with the pure surfactants.

Based on the observations made so far, we can safely state that if the organic phase becomes enriched with surfactant, inverted (i.e., water-in-oil) emulsions can form provided that the organic phase contains water as well. This is an important issue, as the particular system that we are interested to resolve is that of an aqueous phase with surfactant contacting an organic phase without the surfactant. One expects that inverted emulsions would form in the organic phase provided both the surfactant and water diffuse into it from the underneath aqueous phase. This issue was addressed previously by Davies, J. T.; Wiggill[55].

This issue is also relate to a recent interesting study by Yang and Abbott[56] , who reported the growth of water droplets on a hydrophobic surface immersed in water, where water vapor had to pass through a layer of oil following which it could undergo nucleation and growth on the hydrophobic substrate. Hence, there already exists evidence that water can dissolve in, and diffuse through, oil. If such is the case, the results summarized in Figure 3 suggest that inverted emulsion would form in that phase in the presence of surfactant. Enrichment of an organic phase by a surfactant due to diffusion from the aqueous phase is directly detectable in UV spectroscopy (see below). The issue regarding water diffusion is addressed next.

**Evidence of Water in Organic Phase**. One straight forward way to resolve this issue is to measure the amount of water in the organic phase using a standard Karl Fischer test after equilibrating a given amount of this phase with an aqueous salt solution. The results summarized in  **Table S2** (Supplementary Materials) show that all organic solvents that are devoid of water to begin with collect certain amount of water after equilibration. The actual emulsification process is, however, more complex, as when the inverted emulsions are formed, the amount of water would be depleted from the organic phase. However, this loss would be compensated by the continuing transfer of water from the aqueous to the organic phase. To make a realistic assessment of the dynamic evolution of water in the organic phase, we designed an experiment by modifying a strategy that we reported in the past[57].

Previously, we reported an experiment in which a drop of a concentrated aqueous solution of a mixture of sugar (i.e., glucose and sucrose) resembling that of a plant-based nectar was exposed to a humid atmosphere[57]. As such a nectar drop is very hygroscopic, it grows by absorbing moisture from the environment. Here, we performed a similar experiment, in which a nectar drop was immersed in the organic phase by keeping it slightly above the interface of the organic liquid and salt water.

The results summarized in Figure 4 show that the small nectar drop grows at a rate, which, as a first approximation, is nearly independent of the nature of the organic liquid. As the osmolarity of water in the nectar drop continues to decrease due to the dilution of sugar, its growth kinetics slows down and is eventually halted. Let us consider a typical experiment in which we consider a small nectar droplet (0.64 µl) of osmolarity of 7.5, which absorbs water and grows. The final volume of this droplet should be about 2 µl for its osmolarity in the diluted state to be same as that of salt solution ( 1.2 moles/L). This is in the ball part of  what is observed experimentally (Figure 4).

The chemical potential of water in the nectar drop, however, is not only determined by the concentration of sugar but also by the Laplace pressure resulting from its curvature and its interfacial tension. In the beginning, the difference of the osmolarity is the main driving force. However, at a later stage, when the difference of the osmolarity becomes small, the Laplace pressure would control the growth process. As the interfacial tension of the nectar drop is lowered due to the adsorption of the surfactant at the interface of the nectar drop and oil phase, the equilibrium size of the nectar drop with surfactant is expected to be higher than that without it.

The above consideration helps us to understand the fact that at equilibrium, the volume of the droplet with surfactant is higher than that without it.  The higher growth rate of the drop with surfactant is likely due to the fact that when water-in- emulsions continue to form it readily supply water to the organic phase when it is depleted. All these lead to a situation known as "facilitated transport". We postpone a detailed analysis of this interesting phenomenon to future.

We also note what happens when certain polymers are dissolved either in the organic or in the aqueous phase. These results reported in Figure 4 show that the growth of the droplet is arrested, suggesting that that polymers thwart the solubility and/or the diffusion of water from the aueous to the organic phase. This result is corroborated with the the Karl Fisher test (Table S2, Supplementary Materials) that exhibits undetectable amount of water in organic phase in the presence of such additives.

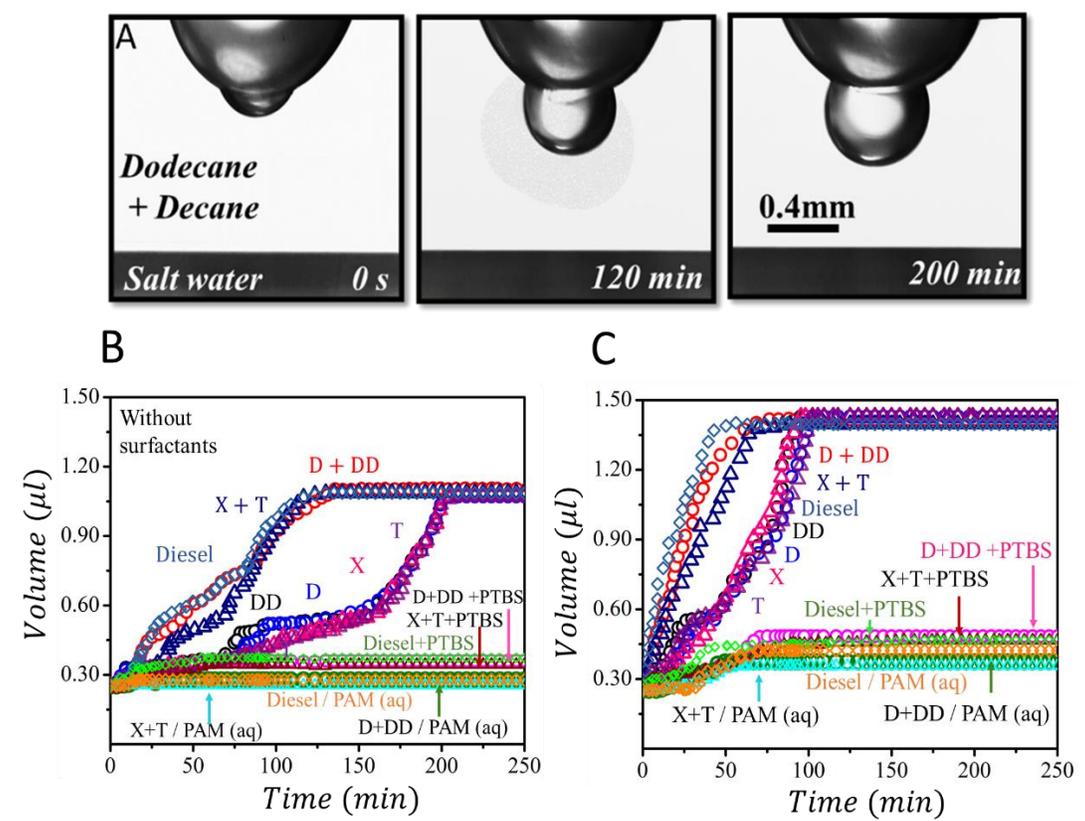

**Figure 4 (A)** A nectar drop (on a glass support) is immersed in an organic phase and placed about 0.7 mm above the interface of the organic phase and 0.6 M salt solution. As water molecules escape from the aueous phase to the organic phase, the nectar drop absorbs them and grows. (B) Growth of nectar drops with and without the presence of surfactant are shown in Alkanes (such as Decane, Dodecane and their mixtures), Aromatics (such as xylenes, Toluene and their mixtures) and Diesel. (C) Same as B, in this case with surfactants dissolved in the aqueous phase. Here, X=xylene, T=toluene, DD=dodecane, D=decane, PTBS= poly-4-tert-butyl styrene, PAM = polyacrylamide (20 ppm) and PTBS 18 ppm in the organic phase.

A short summary of what we learned so far is:

1. Dodecanoic acid interacts cooperatively with Polyoxyethylene (23) Lauryl Ether. This interaction is stronger in the presence of a small amount of water.

2. Remarkably, when these surfactants are pre-dissolved in dodecane and water vapor is infused in that phase, water-in-oil emulsions can form.

3. If surfactants are present in an oil, a small amount of water is needed to form water in oil emulsions. While they can form with either of the anionic or non-ionic surfactants, they are not as stable as those formed when both the surfactants are present together.

At this convenient juncture, we wish to point out that a DPD-MD simulation indicates that oil-in-water emulsion can form in the aqueous phase in the presence of a small amount of oil the aqueous phase. However, the stability of the emulsions formed with the mixed surfactant far greater than those formed with the pure surfactants. While this trend is similar to what we observed with the water-in-oil emulsions, this type of emulsification requiring the diffusion of oil from the organic to the aqueous phase would be a slow process as the solubility of oil in water is very low.

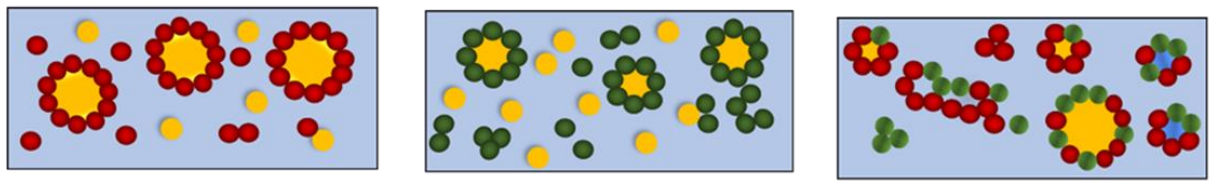

*DDA, $t_-/t_+ = 0.001 - 0.005$*  *Brij-35, $t_-/t_+ = 0.001 - 0.003$*  *DDA+B, $t_-/t_+ = 0.2 - 3.1$*

**Figure 5.** The results of the DPD-MD simulations carried out with 100 ppm of surfactant in aqueous salt solution in the presence of 1000 ppm of dodecane. Stabilities of the different aggregates can be discerned from the ratios of $t_-$ and $t_+$ (see text). The color coding are as follows, Red: DDA, Green: Brij-35, Blue: water, Yellow: Oil)

Some of the fast kinetics of emulsification that we report next indicates a different route to emulsification that involves a catastrophic instability of the interface following which the inverted emulsions formed in the organic phase cross the interface and transform into  oil-in-water emulsions.

**Experiments on spontaneous emulsification and analysis.** In order to proceed further, an experiment was performed, in which a thin layer of organic liquid (2.2 ml) was deposited above a 2.2 ml of 0.6 M salt solution containing a surfactant inside a quartz cuvette (Figure 6). The height of the quartz cuvette was so adjusted inside the spectrometer sample holder that the UV beam passed either above the interface of the salt solution and the organic liquid,  slightly below (1 mm) and even farther below (4 mm) the interface. When the UV beam passed through the cuvette above the interface, it interrogated the sequences of events occuring in the organic phase, i.e. the fate of the surfactants as it diffuses from the aqueous to the organic phase. The UV beam passing through the aqueous phase, similarly,  interrogated the sequences of events occuring in the aqueous phase. One disadvantage of the current experimental procedue stems from the limitation that the organic and aqueous phases could be studied simultaneously. These experiments had to be performed separately in order to collect the information of the events occuring in the three regions, being aware of the possibiliy that there could be some discrepancies in the time courses of the events that we are interested to explore. Nevertheless, with adequate care, these experiments could be performed that provided meaningful picture of the phenomena with sufficient details and acceptable reproducibility.  Although these experiments were  performed with various types of organic liquids, here we present the results obtained with a few selected liquids. Additional results are summarized in Figure 7**.**

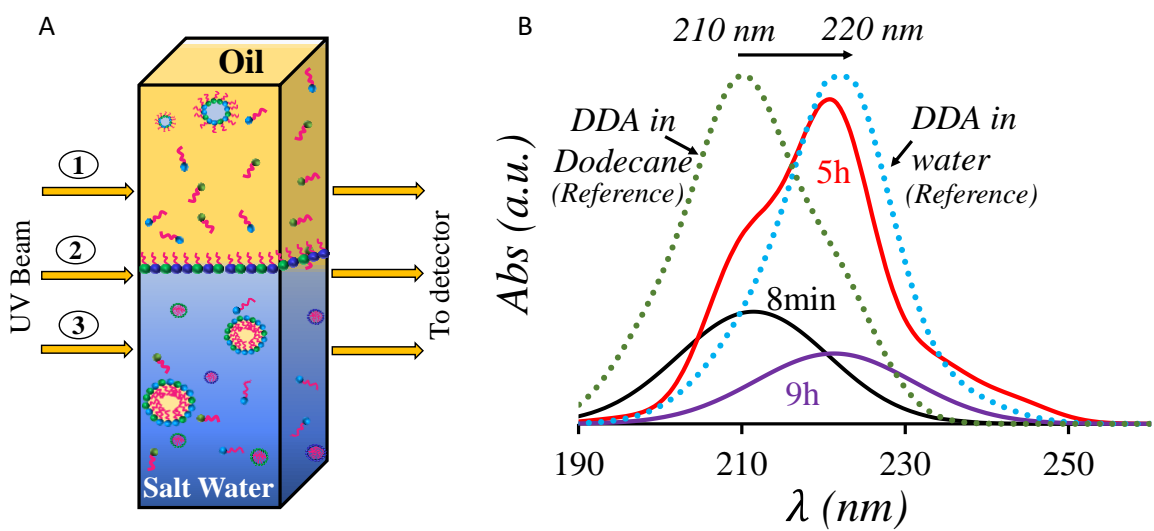

**Figure 6.** (A) Interrogation of an oil phase (dodecane) about 4 mm (as indicated by UV beam in A1) above the interface shows the enrichment of this phase with dodecanoic acid with time. (B) The primary absorption peak is observed at about 210 nm, which is lower than that (210 nm) in water. As time progresses, this peak diminishes with the concomitant increase of a peak close to 220 nm, which is observed when the surfactant is in water. At long time, the primary absorption occurs at 220 nm, which, however, decreases with time. See also Figure 7, below.

**Evidence of Inverted Emulsion in Organic Phase.** The results summarized in Figure 6B illustrate the evolution of the concentration of DDA in an organic phase (dodecane). Initially, the primary absorption peak is observed at 210 nm, which is lower than that (220 nm) of the surfactants dissolved in an aqueous phase. This indicates that DDA, which engages in extended H-bonding in the aqueous phase breaks free when it first enters the organic phase, as evidenced from the observation that the primary absorption peak is found to be same as that of the surfactant in the organic phase in the absence of water. With the progression of time, a peak appears at about 220 nm corresponding to the hydrated state of the surfactant, while the 210 nm peak continues to grow and persists for a certain amount of time. At a longer time, the peak at 210 nm virtually disappears; while the new peak centered around 220 nm spans a broad region (220 nm to about 250 nm), which is similar to what is observed with DDA when it is dispersed in the aqueous phase. This indicates the formation of an extended bonding network in the organic phase. During the formation of this network, both the protonated and the de-protonated carboxylic acid must be participating if we assume that the pH of the water, with which the surfactants nucleate, is close to 7.0 (contrast this value to the pKa of DDA, which is about 7.5). Ultimately, the concentration of the surfactant decreases in the organic phase, presumably due to fusion of the inverted emulsion with the interface and its subsequent re-entry into the aqueous phase in the form of an oil-in water emulsion.

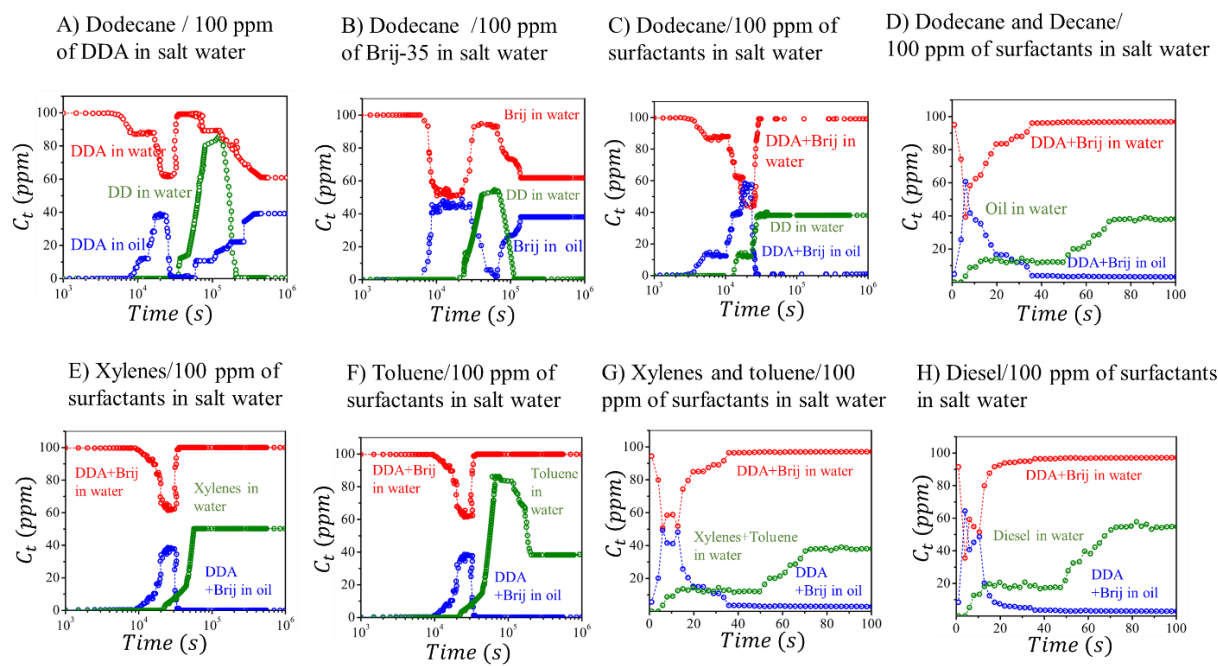

**Figure 7**. Evolution of the concentration of various surfactants and their mixtures in organic and aqueous phases as a function of time. Note the fast dynamics of surfactant transports in mixed solvents (D and G) as well as in Diesel (H), which is composed of various organic solvents.

The overall spectroscopic signature of the surfactant is similar to that shown in Figure 3. Nevertheless, while the total surfactant remains constant in the experiment described in Figure 3, the dynamics of the surfactant in the current experiment results from the combination of the transport across the phases, the transformation from an unhydrated to hydrated states and its subsequent re-entry into the aqueous phase. These kinetic results are plotted over a long period of time in Figure 7, where with the anionic and non-ionic surfactants and their mixtures were dissolved in dodecane, toluene, xylene and in the solutions of decane and dodecane as well as toluene and xylene. In these plots, we also include the evolution of the concentration of essential components in the aqueous phase. As far as the aromatics are concerned, they are UV active; while the alkanes are not. In order to detect the presence of alkane in the aqueous phase, we added a small amont (10 ppm) Triphenylmethane (TPM) in it as spectral reporter. As Triphenylmethane dissolves in the organic phase but is totally insoluble in the aqueous phase, its presence in the aqueous phase via UV spectroscopy signifies the presence of the alkane if it presents itself as a droplet. From a calibration curve of the concentration of this marker in alkane vs UV intensity, the amount of alkane in the aqueous phase could be estimated from the sperctral intensity obtained from a given experiment.

With dodecane above water, we note that the concentration of DDA ($C_{DDA}^{oil}$) initially increases with time as the surfactant diffuses from the salt solution to dodecane. While $C_{DDA}^{oil}$ increases, the concengration of DDA ($C_{DDA}^{Aq}$) decreases in the aqueous phase (Figure 7). Upon reaching the optimal values in the organic and aqueous phases, $C_{DDA}^{oil}$ decreases with a concomittant increase of $C_{DDA}^{Aq}$. This observation coupled with the concomittant increase of concentration of oil in the aqueous phase ($C_{oil}^{Aq}$) indicate that oil-in-water emulsions are formed in the aqueous phase during this time. At the longer time, these emulsions appear to break down as evident from the decrease of $C_{oil}^{Aq}$ while DDA is pseudo-equilibrated in the two phases, when $C_{DDA}^{oil}$ becomes comparable to $C_{DDA}^{Aq}$. While the results reported in Figure 7 correspond to a window of five days, the concentrations continued to evolve and fluctuate beyond beyond that (Figure S6, Suppleentary

Materials). Anlysis of the long term result is, however, compicated as the kinetics at the late stage is is influenced by Ostwald ripening and fusion induced aging of the emulsion droplets. Furthrmore, the large droplets lose the freedom of Brownian motion and tend to accumulate at the interface. We thus stayed away from interpreting these long tme results. Another reason for us to avoid performing these long term measurements systematically is technical, as the spectrophotmeter itself begins to malfunction due to excessive heating of the electronics when it is operated for a very long time.

Experiments carried out with Brij 35 exhibited a behavior similar to DDA, where the evolution of the concentration of the surfactant displays a mirror symmetry in the organic and aqueous phases. While the behavior of the mixed surfactants is similar to that of DDA or brij-35 (Panel 1C and 2C) in the begining, the long term behavior is found to be quite different in that the oil remains in the aqueous phase, quite likely in the form of oil-in-water emulsions. Very similar behavior was also observed with toluene, xylene and other organic liquids (Figure 12). The main conclusion that we can reach from these observations is that neither the anionic nor the non-ionic surfactant forms stable oil-in-water emulsion, whereas stable emulsions are formed with the mixed surfactant, where the coopertivity between the anionic and nonionic surfactants, and the resulting extended H-bond formation appears to play the critical role. See also Figures 2 and 3, and the associated discussions where this issue was highlighted.

A remarkably surprising result is observed when these experiments are performed in a mixture of decane and dodecane or in a mixture of toluene and xylene, which showed a dramatic enhancement of the kinetics of the surfactant transport (Figure 7 (D and G)). Here, the evolutions of the concentration of both the oil and surfactant occur much faster than the pure solvents, in that they reach the pseudo-equilibrium states in less than a minute. Similar results are also found in Diesel (Figure 7H), which , in essence, is a mixture of various organic solvents. Furthermore, when we reconstituted the composition of the diesel using various organic solvents, and performed these experiments, we observed surfactant dynamics that is similar to that of commercial diesel (Figure 7H).

While we do not have a satisfactory explanation for the fast surfactant dynamics observed with the mixed solvents, we do note that the solubilities of the surfactants in the mixted solvents are greater than than those in the pure solvents (Figure 8). While the incresed solubility may play a role in the phenomenon described here, we suspect that its true origin is possibly hidden in the formation of local heterogeneity at the interface thus triggering local Marangoni instabilities. We are encouraged to speculate such a possibility from certain drop spreading experiments, which are not reported here. When a droplet of a pure organic liquid was deposited on the surface of the salt solution containing a surfactant, it spreaded uniformly. However, when a mixture of organic liquids was deposited on the surface of the salt solution, it disintegraed spoontaneously forming multiple thin lense shaped droplets, the edge of which continued to undergo instability and smaller droplets were formed. Postponing further studies on this very interesting topic to future. Now, we report what we learned from the MD-DPD simulations regardig the self assembling behavior of the pure and mixed surfactants in the organic and aqueous phases.

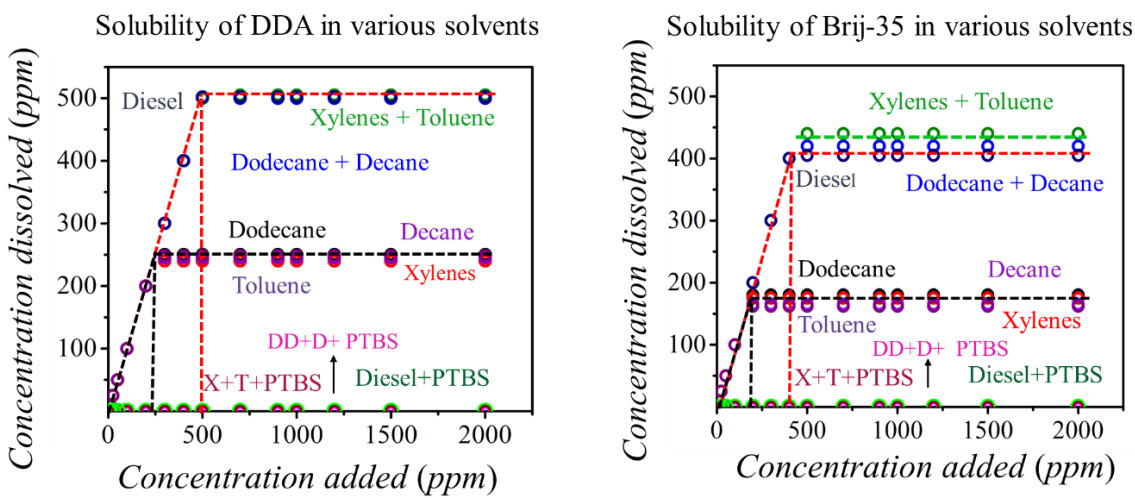

**Figure 8.** Solubility of Dodecanoic acid and Brij-35 in various solvent systems. In all cases, solubility is higher in mixed solvents than in pure solvents. When these experients were carried out with PTBS, it was added in the organic solvent at a concentration of 18 ppm.

The results summarized in Figure 3 show that each of the anionic and the nonionic surfactants can form micelle, although their  stabilities are rather low. On the other hand, when both the surfactants are present, more stable structures are formed in the form of micelles and vesicles, the anatomies of which can be visualized from the plots of the concentrations of water, hydrocarbon, carboxylic acid and ethylene oxide across the diameter of the resultant structures (Figure 9 C).

As expected, the emulsion droplets formed in oil show the presence of water in the core region that is surrounded by a lamela composed of hydrocarbon, ethylene oxide and carboxilic acid groups. On the other hand, the emulsion droplets formed in the aqueus phase show the presence of oil in the core region that is surrounded by a lamela composed of hydrocarbon, ethylene oxide and carboxilic acid groups. Furthermore, for the oil-in-water emulsion, the hydrophilic carboxilic acid and ethylene oxide groups face away from the central hydrocarbon core towards water, the reverse of which is observed with the water-in-oil emulsion. When the mixed surfactants are present in water, both micelles and vesicles appear to form.

The interfacial tensions of the various interfaces could also be estimated using these simulations, which are tabulated in **Tables 2 and S4** (Supplementary Materials). These simulations were carried out using two different grain sizes (0.1 nm³ and 1 nm³) to ensure that the estimated values are independent of the grain size, which is found to be the case. The area per molecule was estimated by counting the numbers of hydrocarbon chains, which are present in both the molecules. The data summarized in Table 2 show that the average interfacial area of the surfactants adsorbed at the flat oil-water interface is close to $0.8\ nm^2$, which is close to the value ($0.9\ nm^2$) obtained from the standard Gibbs' adsorption equation[7]. While a tightly packed hydrocarbon chain occupies an area of about $0.2\ nm^2$, one would expect that the average area per molecule in an emulsion or in a vesicle would be higher than this value as they would be more disordered. On the other hand, the average area occupied by the ethylene glycol group of the non-ionic surfactant should be greater than ($0.6\ nm^2$ ), as obtained [26] from the surface pressure-area isotherm. The average area per molecule, as estimated on the basis of the number of hydrocarbon chains) range from $0.34\ nm^2\ to\ 0.4\ nm^2$ for the emulsions and vesicles (Table 2 and S4).  The fact that this range of values is larger than that $0.2\ \ nm^2$ but less than $0.6\ \ nm^2$ suggests that  the ethylene glycol groups spread above the carboxylic acid groups of the DDA molecules, i.e. the Polyoxyethylene (23) lauryl ether molecules accommodate some of the dodecanoic acid molecules under its umbrella in the self-assembled structures.

A close scrutiny of  all the different cases summarized in **Table 2** reveals, unsurprisingly, that the interfacial tension decreases with the average area per molecule. The interfacial tension of the oil water interface containing 100 ppm of surfactant, as measured by the Du Nuoy ring method, is about 14 mN/m (Table S1) , although a lower value (7.5 mN/m) was obtained with a drop vibration technique. These values can be compared and contrasted with what is obtained in the current simulation. The interfacial tensions of the two emulsions and the vesicles are significantly lower than that of the flat interface, suggesting that the surfactants have organized more tightly in such structures than that at the flat interface, which is a thermodynamic necessity in that the interfacial tension would tend to be as low as possible within the available entropic constraints. What this scenario prompts is the question: what happens when these, presumably, kinetically formed self-assembled units of lower interfacial tension come into contact with the oil-water interface having a higher interfacial tension? If they fuse, the surfactants of these structures would tend to spread at the interface causing a local Marangoni instability. The resulting perturbations could either decay quiescently or violently producing shock waves propagating along the interface. These could have important consequences in the evolution of the subsequent structures. This issue is related to the formation of oil-in-water emulsion at it might happen when an inverted emulsion crosses the interfacial barrier via a singular transformation of one phase into another, which is, otherwise known as phase inversion[40].

**Table 2.** Interfacial tension and molecular area analyzed using MD-DPD simulation using a grain size of 1 $nm^3$.

| Component | Interfcial Tension, $mN/m$ | Total Area $A$ $(nm^2)$ | $N_{DDA}$ | $N_{Brij}$ | $\frac{N_{DDA}}{N_{Brij}}$ | $\frac{A}{N_{DDA}+N_{Brij}}$ $nm^2$ |
|---|---|---|---|---|---|---|
| Flat interface | 8.2 | 10,000 | 5874 | 5425 | 1.12 | 0.88 |
| Oil in Water Emulsion | 1.7 | 493 | 717 | 720 | 1.46 | 0.34 |
| Water in Oil Emulsion | 2.1 | 1319 | 1901 | 1450 | 1.31 | 0.39 |
| Vesicle | 2.3 | 1360 | 2145 | 1530 | 1.21 | 0.37 |

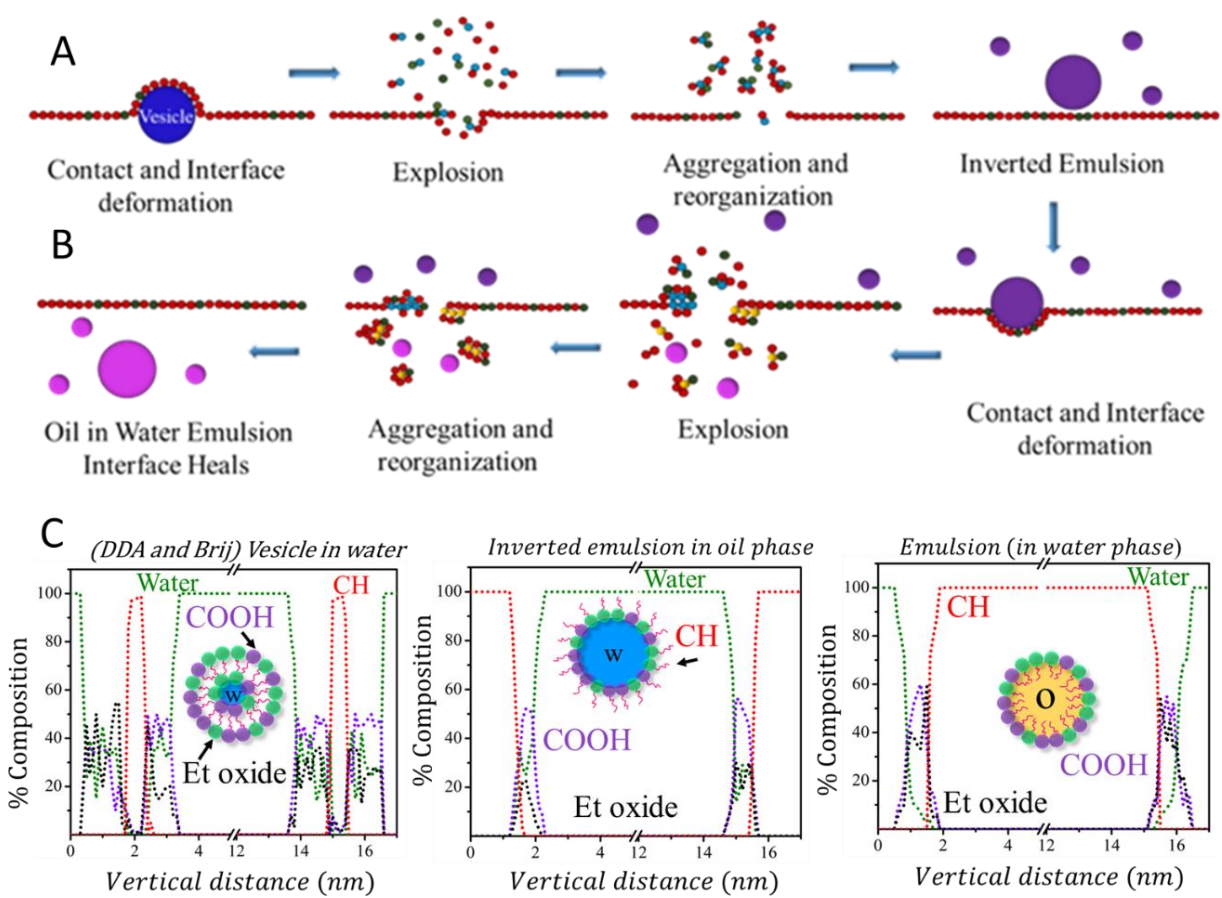

**Figure 9.** Schematics illustrating the formation of water-in oil and oil-in-water emulsions. A) When the surfactants form vesicle in the aqueous phase, it eventually comes into contact with the interface and explodes. The fragments resulting from this explosion cluster together and forms the inverted (i.e. water-in-oil) emulsion. B) When such an inverted emulsion comes into contact with the interface, another explosion occurs, following which the fragments cluster together and transform into oil-in-water emulsions. C) Anatomies of different types of self-assembled structures that are formed either in oil (water-in-oil emusions) or in water (vesicle and oil-in-water emulsions). The color codes as follows, Navy blue: Vesicle, Red: DDA, Green: Brij-35, Blue: water, Yellow: Oil, Purple: inverted emulsion, Pink: Emulsion.

We now address this issue of phase inversion that could occur via the "so called" cusp catastrophe [39-41]. This was accomplished via non-equilibrium DPD-MD simulations, in which an aqueous phase containing either DDA, Brij-35 or their mixture comes into contact with a layer of dodecane. The results summarized in Figure 9A show that the vesicles that are formed in the aqueous phase first come into contact with the interface and deforms it. Following this contact, the interphase region disintegrates rather explosively into various sub-structures, which subsequently cluster together and lead to an inverted emulsion structure in the organic phase. The inverted emulsions eventually come into contact with the interface and undergoes another explosion. The resulting sub-structures then cluster together and form the oil-in-water emulsion, while the interface heals and separates the oil from the aqueous phases. See the supplementary section Figure S10 (A, B and C), which summarizes comparable observations with the micelles formed with pure DDA or pure Brij-35 in the aqueous phase.

This mechanism of formation of water-in-oil emulsion differs from the other, relatively quiescent scenario, which stipulates that the surfactants and water diffuse passively in the aqueous phase, where they self-associate and form nanosized emulsion droplets (e.g. Figure 3). However, the formation of the oil-in-water emulsion from this water-in-oil emulsion via phase inversion should be the same as that shown in Figure 5.

This kind of explosively formed emulsion in the aqueous phase was already hinted in our effort to quantify the amount of oil in the aqueous phase, specifically with the experiments in which a small amount water insoluble marker (Triphenylmethane,TPM) was dissolved in the organic phase . As this marker is totally insoluble in the aqueous phase, its presence there strongly indicates that it does not evolve in the aqueous phase via simple diffusion. The more plausible scenario is that the oil is injected into the aqueous phase as droplet via transient or persistent shake up events due to instabilities occurring at the interface. Such a picture was presented in our earlier publication[26], where we conjectured that the inverted (water-in-oil) emulsions first fuse with the interfacial monolayer following which a curvature reversal (an instability) develops at the interface, thus triggering the phase inversion. Because of the difference of the interfacial tensions between the flat interface and the water-in-oil emulsion, we surmise that local heterogeneous Marangoni instability to play an important role in the explosive phase inversions as depicted in Figure 9 (A and B).

**Diffusivity and interfacial fluctuation.** While we do not have a direct way to interrogate the above-mentioned explosive events at present, we presume that these events should occur intermittently over a long period of time, the signature of which should be present in the fluctuations of the number of emulsion droplets, and thus the concentration of the organic content inside them.

In order to elucidate the above point, UV spectrophotometric measurement was carried out 1 mm below the oil-water interface (Figure 10A) at a rate of 1 acquisition/second. These data summarized in Figure 10B show that the fluctuation of UV absorbance, which starts almost immediately in some cases as the data acquisition begins, whereas in some other cases, the fluctuation begins after a certain lag time. Note that such lag time was evident in the concentration evolutions as shown in Figure 7. Furthermore, the intermittency occurs at fast pace in some cases whereas they sparsely in other cases.

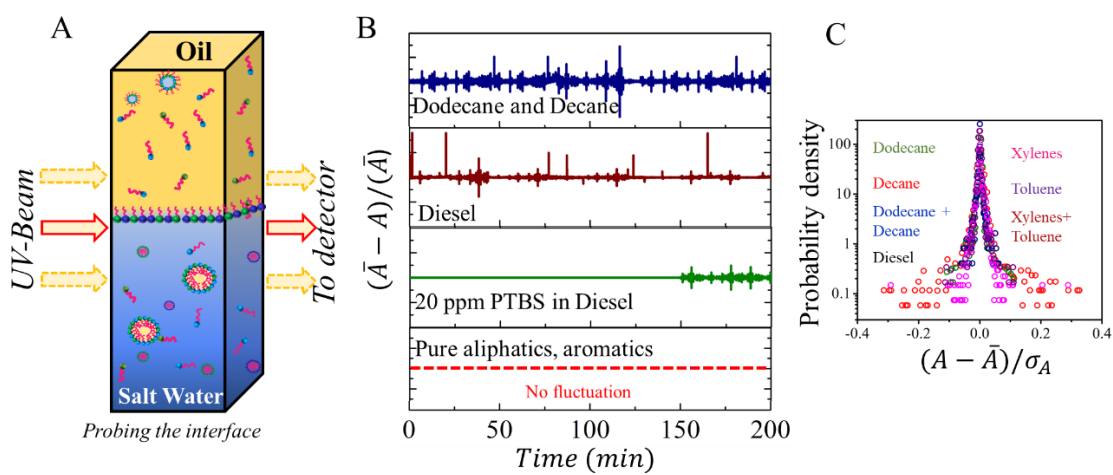

**Figure 10.** **(A)** The region just below the oil-water interface exhibit significant fluctuation of the concentration of the organic phase. **(B)** Different types of fluctuations are observed. In some cases, intermittency is rapid and in some other cases, the intermittency is delayed. **(C)** The probability distribution functions of these fluctuations exhibit non-Gaussian behavior.

For pure liquids, such as alkanes, no such fluctuation of concentration could be observed when surfactant concentration was 100 ppm. However, with the exactly same system, significant fluctuations were observed when the concentration of surfactant increased to about 500 ppm. The power spectral densities of the time series of the concentration fluctuation generally exhibit bluish noise, with the probability distribution function of the fluctuation exhibiting a non-Gaussian behavior. We surmise that they are all related to the burst events resulting from the explosive formation of emulsions as observed in the DPD-MD simulations.

The emulsions produced below the interface should diffuse farther into the aqueous phase. If the concentration is probed at a fixed distance $x$ from the interface, then it should evolve according to equation 1.

$$\frac{C_0 - C(t)}{C_0} = \text{erf}\left(\frac{x}{\sqrt{4Dt}}\right) \tag{1}$$

where, $C_0$ is the concentration at saturation, $C(t)$ is the concentration at a given time $t$, and $D$ is the diffusivity. Equation 1 can also be re-arranged as follows:

$$erf^{-1}(\varphi) = \left(\frac{x}{\sqrt{4Dt}}\right) \tag{2}$$

Were, $\varphi = \frac{C_0 - C(t)}{C^*}$ . Thus, a plot of $erf^{-1}\varphi \; vs \; 1/\sqrt{t}$ should be a straight line from the slope of which, the diffusivity can be estimated. Since there is a lag time ($t_0$) for the diffusion front to reach the region where the concentration is measured, iy makes sense that it the diffusivity be estimated from the slope the plot of $erf^{-1}\varphi \; vs \; \frac{1}{\sqrt{(t - t_0)}}$ in the long-time limit, i.e. when $t \gg t_0$.

In order to estimate the diffusivity in the short time limit, we use the time $t_0$ that is elapsed before the first sign of organic liquid is detected in UV. Taking $\varphi \approx 1$ at $t = t_0$, diffusivity can be estimated from equation (2) using $x^2 = \pi D t_0$ . These diffusivities are normalized by dividing their values with the respective Stokes-Einstein diffusivities: $D_e = k_B T/(6\pi\eta r)$ , where $k_B$ is the Boltzmann constant, $T$ is the temperature, $\eta$ is the viscosity of the aqueous phase. $r$ is the average radius of the droplet, which was estimated[26] using CHDF (capillary hydrodynamic flow chromatography; Figure S9, Supplementary Materials). Since the fluctuation of concentration below the interface encodes the magnitude of the diffusivity of the emulsion droplets, we found it convenient to show the wide range of the value of $D/D_e$ , by plotting it against mean square value of fluctuation $\sigma_A^2$ of UV absorbance (which is a measure of $\sigma_C^2$) as obtained from Figure 10B.

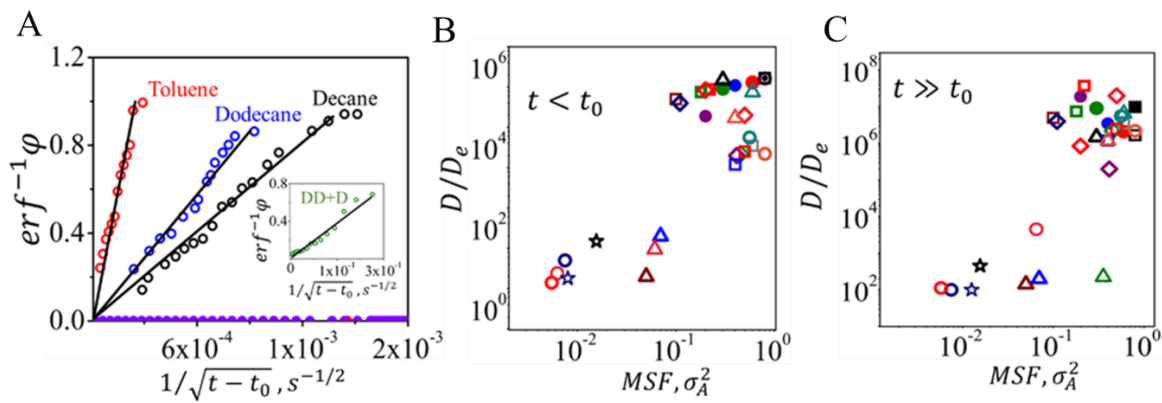

**Figure 11.** A) $erf^{-1}(\varphi)$ is plotted against $1/\sqrt{(t - t_0)}$ for a few representative cases, where $t_0$ is the dead time following which the concentration begins to rise, $\varphi = [(C_0 - C(t))/C_0]$, $C_0$ being the saturated concentration, at long time. The slope of this curve gives the estimate of diffusivity at short (B) and long times (C). In B and C, The ratio of the estimated diffusivity and the Stokes-Einstein diffusivity De is plotted against the mean square fluctuation of UV absorbance (hence concentration). See also Figure 10B. The detailed designations of the various cases can be found in Tables S5a and S5b (Supplementary Materials).

Figures 11 (B and C) show that the effective diffusivities of emulsions produced in the aqueous phase exhibit a wide range of $D/D_e$ values that, in general, increase with $\sigma_A^2$. Had the diffusion of droplets been governed entirely by usual thermal noise, $D/D_e$ would be close to unity. There are only few cases, which present this type of value [Table S5a], for which $\sigma_A^2$ values were undetectably small that they could not be potted along with the other data in Figure 11 . However, in most cases $D$ is several orders of magnitude higher then $D_e$. The fact that $D \gg D_e$ indicates that the droplets acquire certain amount of additional kinetic energy as soon as they were formed in the aqueous phase below the interface. We conjectured such a possibility previously; however, the current studies clearly suggest that such a fast diffusion could be controlled by the non-thermal fluctuations that result from the explosive events reported in Figure 9 (A and B).

**Consolidation of the important points.**

The research presented here is a modest contribution to the saga that has continued over a century to understand the mechanism of spontaneous emulsification. Interrogating the transfer of surfactants and oil (or the organic liquid) using UV spectroscopy helped develop an intuitive picture of the relevant processes, whereas the non-equilibrium DPD-MD simulations provided a complementary insight needed to understand the local processes. Some of the main observations made in these studies are as follows.

1. When either an anionic or a non-ionic surfactant containing aqueous phase contacts an organic liquid, the surfactant and water can diffuse into the organic phase, where inverted (water-in-oil) emulsions are formed.

2. A parallel mechanism for this inverted emulsification is that the micelles or the vesicles present in the aqueous phase fuse with the interface explosively, following which the ensuing fragments self-assemble and form water-in-oil emulsions.

3. In both cases ( 1 and 2 above) , the inverted emulsions fuse with the interface rather explosively, but intermittently. The resultant fragments then turn into an oil-in-water emulsion.

This current picture is related to, yet it differs from, what we conjectured previously, in that the inverted (water-in-oil) emulsions was thought to fuse with the interface, following which a curvature reversal induced instability develops at the interface. In this work, the non-equilibrium MD-DPD simulations identify the intermediate state, which is the reminiscent of what was called cusp catastrophe by Dickinson[40].

We feel that more future work is warranted that would couple a detailed analysis of the thermodynamic and kinetic processes in unravelling new and rich physics underlying such types singular transformation of one type of phase into another.

A remarkable observation is that the transport and self-association dynamics of the surfactants is faster in the mixed solvents than that in the pure solvents. Whenever such fast events occur, the concentration fluctuation near the interface is amplified, concomitantly. A question here is whether the origin of this phenomenon is thermodynamic or kinetic? A partial answer of this question comes from the water infusion induced emulsification studies of the type presented in **Figure 3**. In these experiments, surfactant is already dissolved in the organic phase before passing water vapor through it. Thus, any emulsification, if it occurs, would here be de-coupled from any kinetic processes related to the transfer of water and surfactant from the aqueous to the organic phase. The results summarized in Figure 7 show that emulsification kinetics, as evidenced from the shift of the UV spectra, is always immensely faster in mixed solvents than in pure solvents, just is the case with both surfactant and water migrating into the organic phase from the aqueous phase below it. Diesel is essentially a mixture of various aliphatic and aromatic hydrocarbons, which thus falls in the same group as mixed solvents. Enhanced solubilities (Figure 8) of the surfactants in the mixed solvent, although the underneath cause of which remain unclear, suggest that part of the effect is thermodynamic in origin, we cannot rule out the possibility that water vapor may engender an instability at the vapor-liquid interface that being the cause of the fast kinetics observed here.

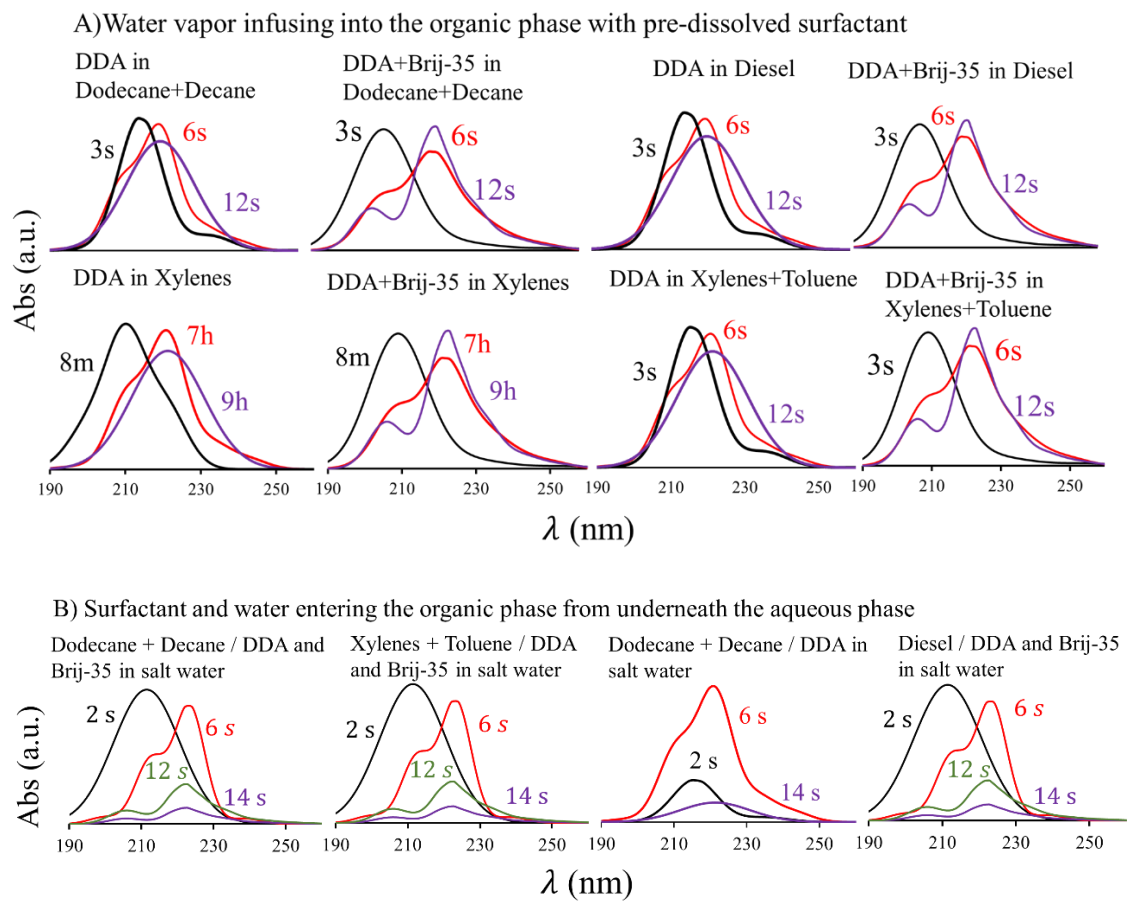

**Figure 12**. Bathochromic (red) evolutions of the UV spectra of Dodecanoic acid, Brij-35 and their mixture (50:50) in various organic solvents and their mixtures. See also Figures 3G and 6B. Two points can be noted here. All the Bathochromic shifts occur much faster in mixed aliphatic or organic solvents compared to any pure solvent.

Another remarkable observation is that the emulsification is inhibited when either a high molecular weight hydrophilic polymer is present in the aqueous phase or when a high molecular weight hydrophobic polymer is present in the organic phase. Some of the experimental results performed with PTBS are summarized below.

1. The water infusion experiments of the type presented in Figure 13 (A) show that the dodecanoic acid remains unhydrated when PTBS is pre-dissolved in the organic liquid. Note that similar results are also obtained in experiments, where surfactant and water have to migrate into the organic phase from the aqueous phase. While the latter experiment is dependent on the solubility and the mass transfer kinetics of water from the aqueous to the organic phase, which are drastically arrested by PTBS (Figure 13 (B)), emulsification in the water infusion experiment is decoupled from the issue of surfactant transport.

2. No fluctuation of concentration of oil under the interface is evident in the types of experiments reported either (Figure 10 (B)). This is, however, possibly an effect ensuing from the causes related to suppressed diffusivity and solubility of DDA in the organic liquid in the presence of PTBS.

3. After a very long time, the surfactants are only mildly depleted from the aqueous phase with a concomitant increase of the surfactants in the organic phase and a slight increase of oil in the aqueous phase (Figure 13 (C and D)). All these concentrations are eventually re-equilibrated following which DDA and oil virtually disappear from the aqueous phase.

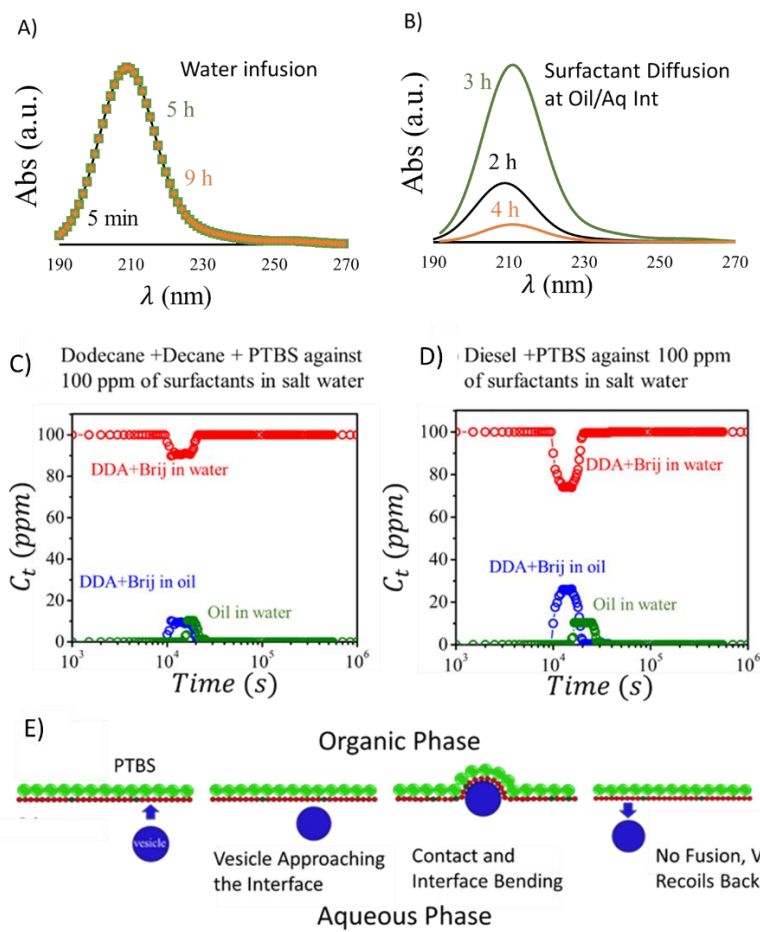

**Figure 13.** A and B: Evolution of the concentration of a mixture of DDA and Brij-35 in organic and aqueous phases as a function of time. C) Schematics illustrating that PTBS inhibits formation of the inverted emulsion . The color coding are as follows, Navy blue: Vesicle, Red: DDA, Green: Brij-35, Fluorescent green: PTBS) 18 ppm PTBS

4. Finally, an MD-DPD simulation carried out with a small amount of PTBS (18 ppm) pre-existing in oil show that the polymer adsorbs at the interface from the oil side [60]. A vesicle formed in the aqueous phase deforms the interface somewhat, without undergoing fusion. The vesicle, in turn, is recoiled back into the aqueous phase.

## Conclusions

Spontaneous emulsification occurs when the pre-formed water-in-oil emulsions cross the oil-water interface by undergoing phase inversion. The water-in-oil emulsions can form in one of two ways. The first path to its formation is that the surfactant and water diffuse in the organic phase and self-assemble there. The second path to its formation is the fusion of the micelles and vesicles, which are formed in the aqueous phase, with the oil-water interface. This process leads to explosive fragmentation of the interface locally, before they self-assemble to form the water-in-oil emulsion. What is crucial is that both water and surfactant diffuse into the organic phase from the aqueous phase. If this diffusion and the fusion events occurring at the interface can be arrested, spontaneous emulsification can be thwarted.

## Acknowledgements

Financial support from The Strategic Environmental Research and Development Program (Grant No. WP18-1074, managed by Robin Nissan and Braxton Lewis) is gratefully acknowledged. We also thank Professors El Aasser and Steve Regen for donating various equipments to our laboratory, which were crucial for the studies described here.

## Supporting Information

Text describing experimental details including the materials and the analytical test procedures along with six Figures and three movie clips are available free of charge via the internet at http://pubs.acs.org.

# Supplementary Materials

## Spontaneous Emulsification: Elucidation of the Local Processes

Monicka Kullappan[1,3], Wes Patel[1] and Manoj K. Chaudhury[1,2]

[1]Department of Chemical and Biomolecular Engineering

[2]Department of Materials Science and Engineering, Lehigh University

Bethlehem, Pennsylvania 18015, United States

Number of Pages: 22

**Experimental Methods**

**Materials** Dodecane (reagent plus), Heptane, Toluene, Benzene, Styrene, Xylenes (HPLC grade), Dodecanoic acid (Analytical Standards, 99% pure) and Brij 35 (Analytical Standards, 99% pure) were purchased from Sigma Aldrich, USA. Decane (HPLC grade) and Octane (HPLC grade) were purchased from Fisher scientific, USA. Dieseectarl oil (with sulfur < 50 ppm) was purchased from a local Shell gas station, which was purified as discussed below. Poly (4-tertbutylstyrene) (MW: $1.5 \times 10^6$ Da, 99% pure) was purchased from Polymer source Inc., Canada. Each of the solvents was purified by passing it through a column of activated alumina three times. The activated alumina (AA400G 28x48) used in the column chromatography was purchased from Delta Adsorbents, USA. All the purified samples were stored in airtight glass containers inside a dust free cabinet, unless it is used immediately after purification. D-Glucose (Anhydrous, analytical Standards, 99% pure) was purchased from Thermo Fisher, USA and D-Fructose (Anhydrous, analytical Standards, 99% pure) was purchased from Alfa Aesar, USA. Water used in these experiments was distilled and de-ionized (resistance: 18.0 MΩ), which had a turbidity of 0.1 NTU (Nephelometric turbidity units).

**Analytical and process instruments** The Spectroscopic studies were performed with the NanoDrop-oneC UV-Visible spectrophotometer, purchased from Thermo Fisher Scientific. The droplet size distribution of the emulsion was carried out with a capillary hydrodynamic fractionator (CHDF-2000, MATEC Applied Sciences), after filtering it through a 1.5 $\mu m$ Whatman glass microfiber filter (934- AH, Fisher scientific, USA) before the CHDF analysis (Note: Filtration step is added to avoid any cross contamination in the system; the droplet size distribution did not depend on it). The interfacial tensions (IFT) were measured using the Fisher Autotensiomat surface tension analyzer (Model – 215). IFT measured for various organic phases against salt solution with and without surfactants are reported in Figure S3. The water was distilled using a Durastill Automatic distiller (Model 3040U), following which it was deionized using Barnstead E-Pure deionizer from Thermo Fisher, USA. . The ultrasound-probe sonicator (Sonifier, S-450 (20 kHz)) was used in pulsed mode for the homogenization of the surfactant solutions and dispersion of polymers into the oil. This unit was purchased from Fisher scientific, USA. The viscosity of oil was measured with a cannon U-tube capillary viscometer (H-472). The glass vials and flaks were used for storing the emulsion samples were purchased from Fisher scientific, USA. Vials used as received. Flasks were cleaned with a standard piranha solution followed by rinsing with copious amount of distilled water.

**Spectroscopic studies.** The UV spectroscopic studies were carried out with two types of quartz cuvettes, one of which was a flow cell type cuvette with a side arm, which was used for the water vapor infusion studies (Figure S1). In order to study the role of hydration of the bathochromic shift of a surfactant, it was dissolved in an organic solvent inside a conical flask. Water vapor saturated air (produced by a FTIR dry gas generator from American Laboratory, Thermo Fisher, USA) was bubbled through the organic liquid in the conical flask at a rate of 62 bubbles/min (bubble diameter

– 2-3 mm) while the liquid was stirred continuously with a magnetic stirrer (100 rpm). This liquid was passed (~2 ml/min) through the quartz cuvette [FireflySci's macro spectrophotometer flow through cell (3 mL capacity) with detachable inlet/outlet quartz tubes, Fisher scientific, USA] using a tygon tube that connected the cuvette and the liquid in the conical flask with the help of a Peristaltic pump (Masterflex L/S Easy-Load II and economy drive, Cole-Parmer)

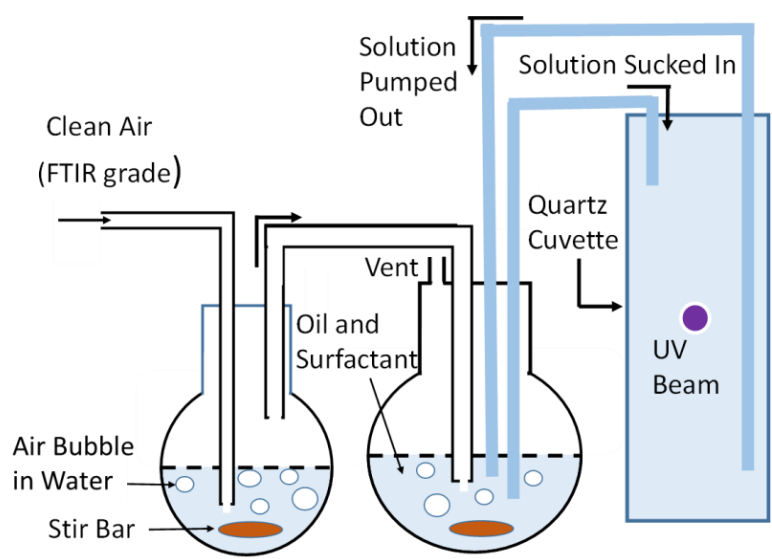

**Figure S1.** Schematic describing how the surfactant dissolved in an organic solvent is hydrated by passing water vapor through it.

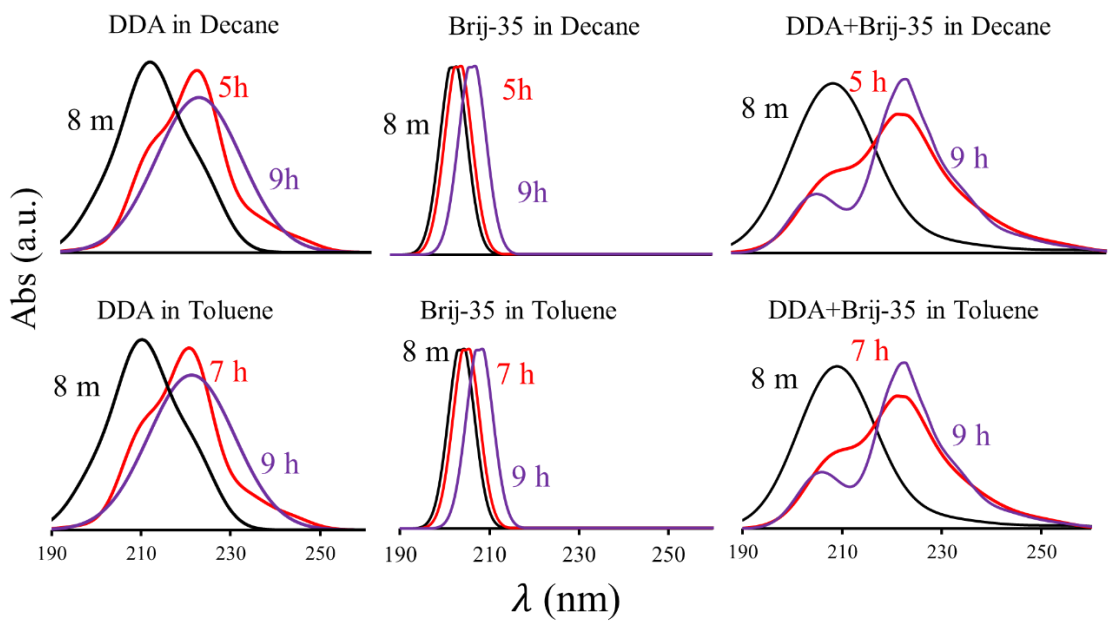

**Figure S2.** Bathochromic (red) evolutions of the UV spectra of Dodecanoic acid, Brij-35 and their mixture (50:50) in various organic solvents and their mixtures as Water vapor infusing into the organic phase with pre-dissolved solvent.

The other quartz cuvette (LAB4US UV Quartz Cuvettes, 10 mm path length from fisher scientific, USA) was used to study the evolution of the concentration of the surfactant as it diffused from the aqueous to the organic phase as well as the evolution of the concentration of oil and surfactant in the aqueous phase. An organic liquid (2.2 ml) was gently deposited above 2.2 ml of 0.6 M salt solution inside the cuvette (Figure 5). The height of the quartz cuvette was so adjusted inside the spectrometer sample holder such that the UV beam passed either above the interface of the two liquids, as shown in the self-explanatory schematic (Figure 9). The UV beam passing through the organic phase monitored the concentration of surfactant as it diffused from the aqueous to the organic phase.

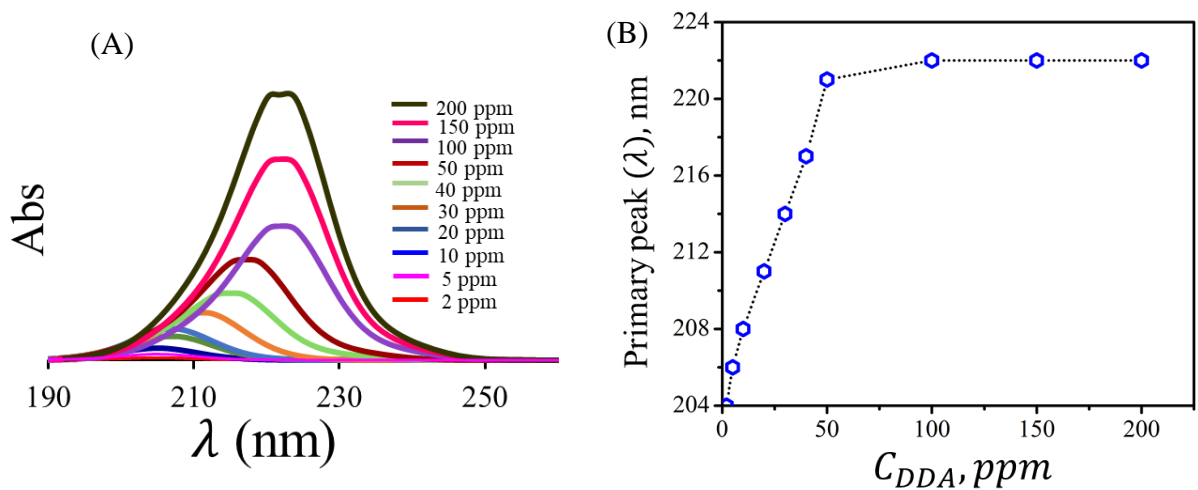

**Figure S3.** (A) UV absorption spectra of dodecanoic acid in 0.6 M salt water as a function of increasing concentration at pH 7. For very low concentration of DDA, the peak maximum appears at 205 nm, which shifts to higher wavelength as the concentration of DDA increases. (B) Shows the plot of primary peak position as a function of increasing concentration of DDA in the aqueous phase at pH 7, which shows a critical concentration (~ 30 ppm) beyond which the spectrum remains unchanged.

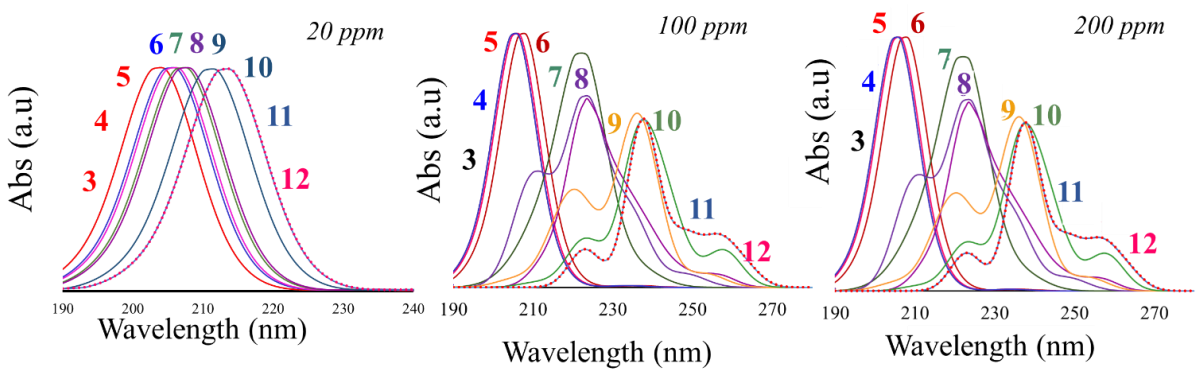

**Figure S4.** UV absorption spectra of dodecanoic acid in 0.6 M salt water as a function of pH for three concentrations of DDA. As the pH becomes more basic, the primary absorption peak exhibits a red shift.

The beam passing through the aqueous phase monitored the concentrations of the surfactant the organic liquid that was emulsified. In order to measure the fluctuation of the concentration of surface below the interface, the height was re- adjusted with the help of Teflon spacers in the holding cell. The cuvette was initially loaded with surfactant solution containing a surfactant of pre-selected concentration, following which, an organic liquid  was added carefully over the aqueous phase so as to avoid undesired agitation. This loaded cuvette placed inside the sample holder of the spectrophotometer was then closed with a Teflon cap and spacers as needed. The kinetic measurements were initiated using a customized kinetic program that could record and store the spectroscopic data at 1 sec interval.

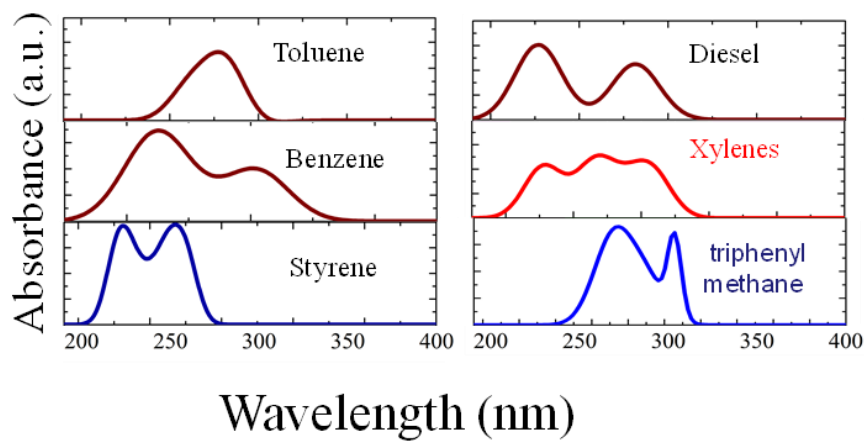

**Figure S5.** UV spectra of the organic liquids used in this experiment. Triphenylmethane is an additive in alkanes, as the latter, itself, is not UV active.

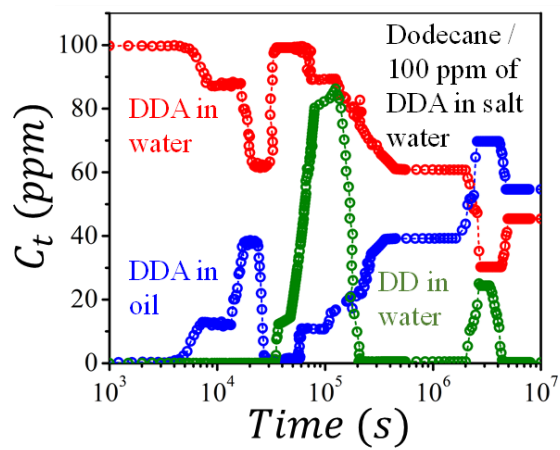

**Figure S6**. Evolution of the concentration of surfactant in dodecane and salt water as a function of time. This Figure represents the data presented in Figure 6, but for longer time.

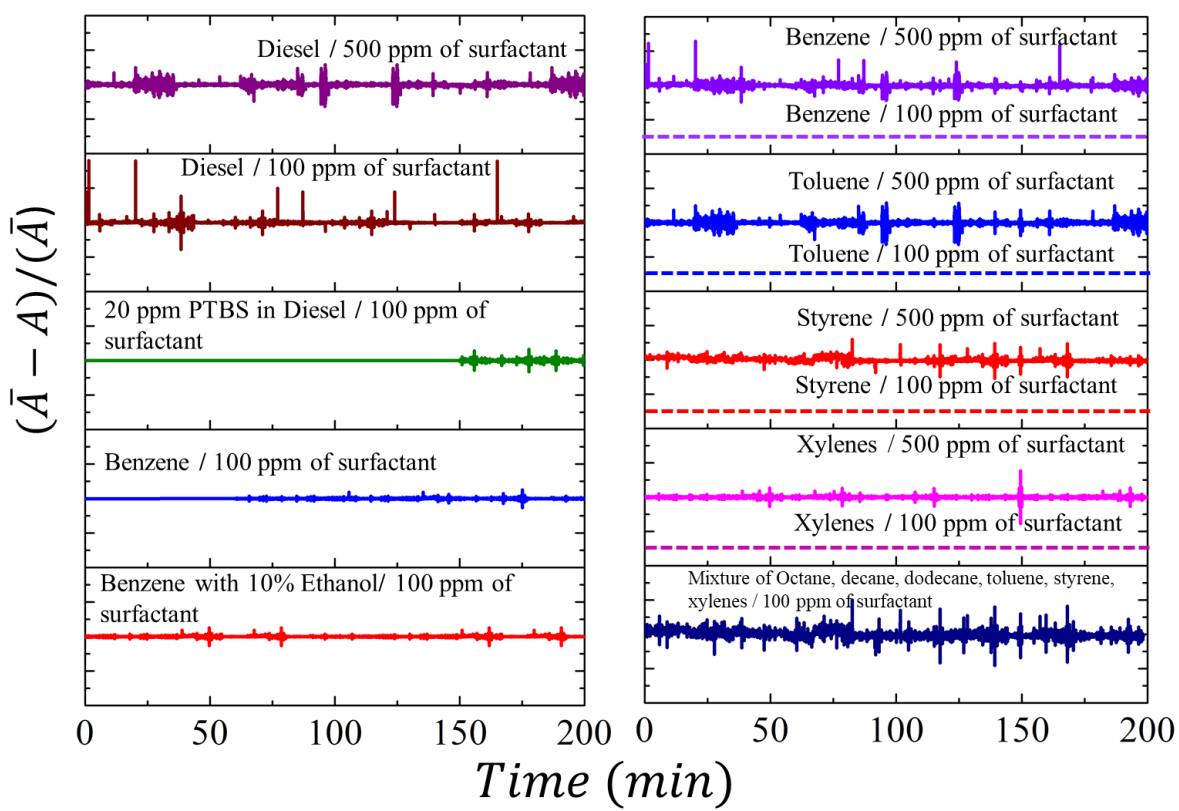

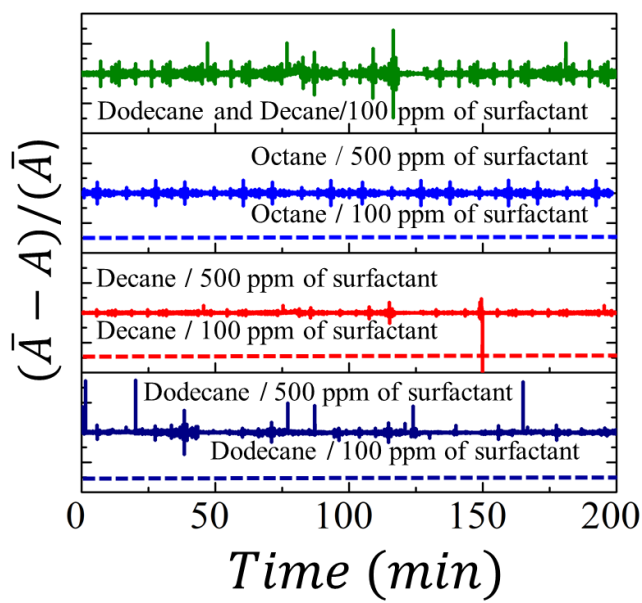

**Figure S7.** Fluctuation of the concentration of the organic phase below the oil-water interface. These results are similar to those reported in Figure 9, but for more different cases.

**Measurement of Interfacial tension using Du Nouy Ring** The interfacial tensions of the oil-water interfaces were measured (23 $^0$C) using the standard Du Nouy ring (Fisher Autotensiomat surface tension analyzer) following standard methods, which we described in reference[26].

**Table S1.** Interfacial Tensions of various liquids against 0.6 M salt aqueous solution with and without the addition of surfactants measured by Du-Nouy ring method.

| Organic solvent | Interfacial tension Without surfactant (mN/m) | Interfacial tension With DDA (mN/m) | Interfacial tension With Brij-35 (mN/m) | Interfacial tension With DDA+Brij-35 (mN/m) |
|---|---|---|---|---|
| Dodecane | 52.0±0.12 | 10.3±0.02 | 15.0±0.30 | 14.2±0.10 |
| Decane | 51.6±0.10 | 13.1±0.01 | 15.8±0.08 | 14.1±0.04 |
| Dodecane+Decane | 51.8±0.01 | 13.0±0.02 | 14.8±0.30 | 14.0±0.01 |
| Octane | 50.8±0.08 | 11.5±0.02 | 14.5±0.25 | 13.8±0.10 |
| Cyclohexane | 51.0±0.11 | 10.8±0.05 | 15.1±0.04 | 14.0±0.10 |
| Benzene | 35.0±0.10 | 9.2±0.01 | 10.2±0.08 | 9.8±0.03 |
| Toluene | 36.0±0.20 | 9.5±0.08 | 11.2 ± 0.05 | 10.0±0.22 |
| Styrene | 37.0±0.10 | 8.9 ± 0.09 | 11.4 ± 0.07 | 10.2±0.23 |
| Xylenes | 36.4±0.10 | 8.5±0.11 | 11.2 ± 0.10 | 10.1±0.21 |
| Xylenes+Toluene | 36.5±0.32 | 10.0±0.04 | 10.4±0.01 | 10.1±0.15 |
| Diesel | 48.0±0.20 | 10.0±0.06 | 12.6±0.24 | 11.8±0.12 |
| Simulated Diesel | 50.2±0.02 | 9.8±0.05 | 12.2±0.07 | 11.2±0.06 |

**Measurement of Viscosity** A standard method, ASTM D445-88, was used to measure the viscosity of the oils used in this study. The dynamic viscosity of the dodecane, decane, octane, xylenes, styrene, toluene, benzene, and diesel were as follows: 1.34, 1.33, 1.46, 1.44, 1.42, 1.42, 1.41 and 1.84 mPa-s respectively. On the addition of 18 ppm of PTBS to various organic solvents, the resulting dynamic viscosities were obtained as 1.35, 1.33, 1.42 and 1.88 mPa-s respectively. From these dynamic viscosity values, it is concluded that there is no significant change in the viscosities of the liquids with the addition of hydrophobic polymers to the oil.

**Study of Spontaneous Emulsification.** The spontaneous emulsification experiment is depicted in Figure S8, a 250 mL conical flask was fixed over the vibration isolation table by means of a metallic stand following which a 100 mL of the salt solution (with/without surfactant) was loaded into the conical flask. This setup was sealed with a Y – shaped glass stopper with two open ports (1.5 cm diameter), which were closed with rubber septum stoppers (1.7 cm diameter). A needle was inserted into one of the septa and a 0.2 mm (I.D.), whereas a Teflon capillary tube was inserted into the other septum stopper. The needle inserted into the first stopper acts as a vent to the atmosphere. One open end of the Teflon tubing was inserted 2 cm below the surface of water This terminal end of this tube was connected to a polyvinyl stopper and a needle assembly, through which the liquid sample from inside the conical flask was siphoned out drop by drop a collected at a rate of ~1 mL/min. After making sure that the Teflon tube was fully filled with the aqueous phase, and the disturbance of the surface subsided, 30 mL of oil (with/without additive) was gently poured over it. Before collecting the emulsion, some amount of the salinized water that pre-filled the tube was discarded, following which, the samples were collected at pre-set intervals.

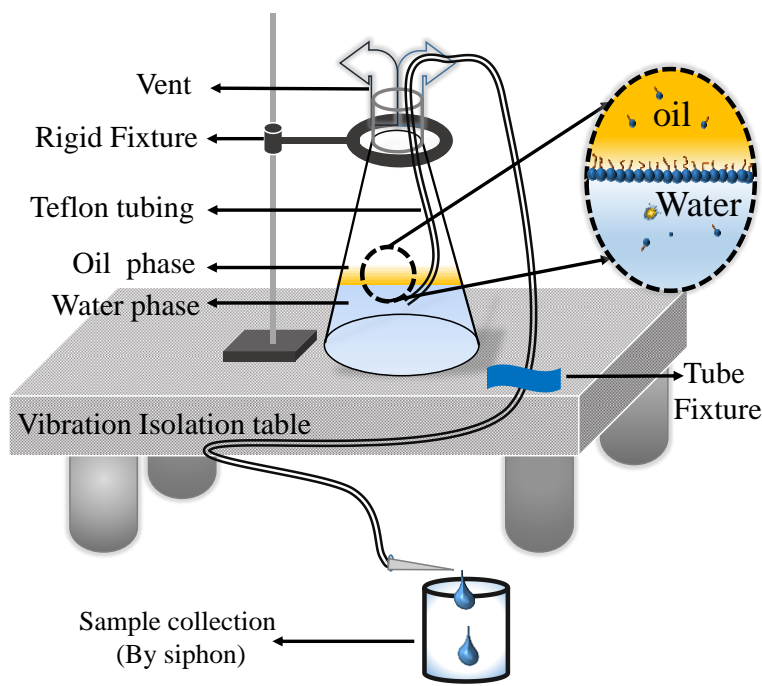

Vent
Rigid Fixture
Teflon tubing
Oil phase
Water phase

oil
Water

Vibration Isolation table

Tube Fixture

Sample collection
(By siphon)

**Figure S8**. A conical flask containing the salinized water and the organic liquid was placed on a vibration isolation table to minimize the effects of ground vibration. The aqueous phase below the organic liquid was collected by a slow siphonic process in which the aqueous phase was collected drop by drop at a rate of about 1 mL/min. Examination of this liquid using CHDF showed the presence of sub-microscopic emulsion droplets of various sizes, but all were less than 600 nm. An optial microscopic image showed the presence of such nano-droplets; however, its size could not be determined by this method as it was less than the optical resolution. After collecting the aqueous phase in a glass vial and examining it in CHDF in subsequent days showed that the droplet size distribution shifted to higher values either via Ostwald ripening and/or fusion.

**Capillary hydrodynamic fractionation** The submicroscopic emulsion droplets were analyzed using the method of capillary hydrodynamic fractionation after the emulsion was passed through a glass microfiber filter with a pore size of $1.5 \ \mu m$. $50 \ \mu L$ of the filtered sample were injected through the sample port to an eluent stream followed by the equal amount of marker solution. A standard retention time of 10 min was used for analyzing all the samples reported in this study.

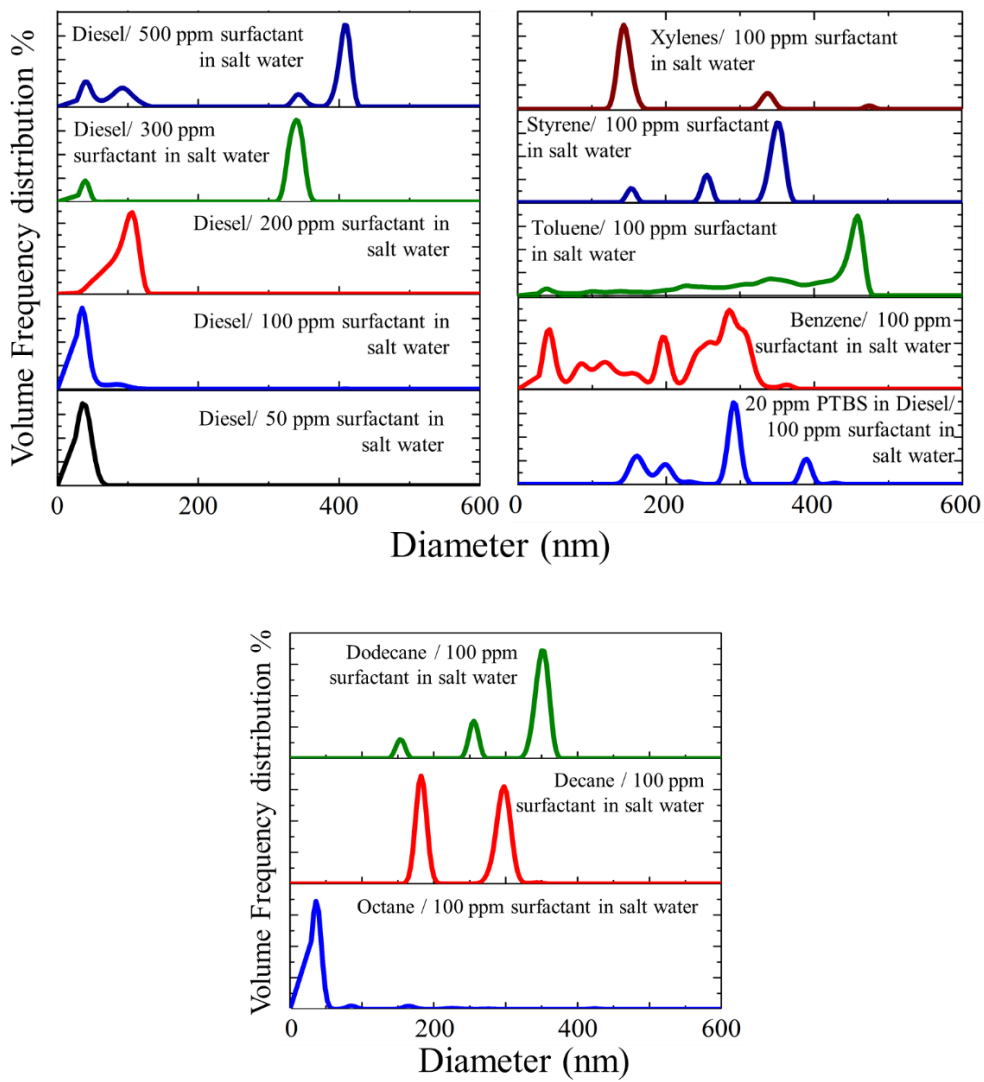

**Figure S9.** CHDF spectra of the spontaneously formed emulsion droplets in the aqueous phase (0.6 M salt solution).

**Solubility of Surfactants in different solvents.** Known quantities of the surfactants were dissolved in the solvents individually in increasing concentration, until further dissolution is thwarted. This surfactant saturated oil was sealed and shaken vigorously in an orbital shaker for 5 min and then stored in the dry air purged environment for 15h. This solvent mixture was then filtered through a 1-micron pore size, Whatman filter and then analyzed in UV-Vis spectroscopy. The calibration plot for the concentration of the surfactant as a function of absorbance value in the respective solvent are described in our previous publication[49]. The point of intersection of the slopes of the region where there is change in the absorbance to the saturated region gives the direct measure of the equilibrium solubility of the surfactant mixture in the corresponding solvent.

**Evidence of Water in Organic Phase**. The concentration of water in the organic solvent which is in contact with the aqueous phase was estimated by Karl Fisher water content analysis. The following protocol were used for analyzing various organic solvents: Each organic solvent (such as Decane, Dodecane, Diesel, Xylenes and Toluene) were individually purified by passing it thrice through activated alumina columns. These purified solvents were initially analyzed for estimating the concentration of water in it by Karl Fisher analyzer (Mettler Toledo, automatic titrator, Robertlit Laboratories, USA). At this stage, the studies on equilibrium water content in each of the organic solvent which are in contact with aqueous phase were systematically conducted (as experimental system listed in Table S1) after storing it for 24 h in isolation chambers to prevent cross contamination. In-order to estimate the water concentration in the organic phase after equilibration with the aqueous solution, a thoroughly cleaned conical flask (50 mL) was initially filled with 20 mL of aqueous phase (0.6 M salt water) in the presence (50 ppm of DDA and 50 ppm of Brij-35) and absence of surfactants (with and without addition of 20 ppm of PAM), following which, 20 mL of purified organic solvent (With and Without 18 ppm of PTBS pre-dissolved in it) was gently deposited over the aqueous phase using a Teflon tube (Dia = 1.5 mm). This setup was carefully made on a vibration isolation table. At this stage, the Teflon tube which is immersed in the organic phase whose other end is suspended freely in the atmosphere (with closed valve) was carefully siphoned to collect the organic solvent at the end of 24 h. All the above collected samples were tested using a standard Karl Fisher (KF) moisture/water content analysis technique (results tabulated in Table S1).

**Table S2.** Karl-Fischer analysis of various organic liquids equilibrated against 0.6 M salt water solution in the presence and absence of surfactants pre-dissolved in the aqueous phase. In all cases surfactant is a mixture of a 50ppm amount of a non-ionic surfactant (Brij 35) and Dodecanoic acid.

| System | Concentration of water (ppm) |
|---|---|
| Dodecane | $< 0.01 \pm 0.001$ |
| Decane | $< 0.01 \pm 0.002$ |
| Dodecane+Decane | $< 0.01 \pm 0.001$ |
| Dodecane/ salt water | $23 \pm 0.12$ |
| Dodecane/Surfactant in salt water | $28 \pm 0.22$ |
| Decane/ salt water | $20 \pm 0.52$ |
| Decane/ Surfactant in salt water | $28 \pm 0.35$ |
| Dodecane + Decane / Salt water | $72 \pm 0.15$ |
| Doecane+Decane/ Surfactant in salt water | $89 \pm 0.85$ |
| Doecane+Decane+18 ppm PTBS/salt water | $< 0.01 \pm 0.001$ |
| Doecane+Decane+18 ppm PTBS/ Surfactant in salt water | $< 0.01 \pm 0.001$ |
| Dodecane+Decane/20 ppm PAM in salt water | $< 0.01 \pm 0.003$ |
| Dodecane+Decane/20 ppm PAM in Surfactant in salt water | $< 0.01 \pm 0.001$ |
| Diesel | $< 0.01 \pm 0.002$ |
| Diesel+20 ppm PTBS | $< 0.01 \pm 0.001$ |
| Diesel/salt water | $102 \pm 2.56$ |
| Diesel/ Surfactant in salt water | $193 \pm 3.15$ |
| Diesel+20 ppm PTBS/salt water | $12 \pm 0.1$ |
| Diesel+20 ppm PTBS/ Surfactant in salt water | $18 \pm 0.12$ |
| Diesel/20 ppm PAM in salt water | $25 \pm 0.07$ |
| Diesel/20 ppm PAM in Surfactant in salt water | $30 \pm 0.22$ |
| Xylenes | $< 0.01 \pm 0.001$ |
| Toluene | $< 0.01 \pm 0.001$ |
| Xylenes+Toluene | $< 0.01 \pm 0.002$ |
| xylenes+toluene+20ppm PTBS | $< 0.01 \pm 0.001$ |
| Xylenes/salt water | $32 \pm 0.21$ |
| Xylenes/ Surfactant in salt water | $48 \pm 0.32$ |
| Toluene/salt water | $15 \pm 0.24$ |
| Toluene/surf in salt water | $23 \pm 0.18$ |
| Xylenes+Toluene/salt water | $38 \pm 0.15$ |
| Xylenes+Toluene/ Surfactant in salt water | $50 \pm 0.11$ |
| Xylenes+Toluene+20 ppm PTBS / salt water | $< 0.01 \pm 0.001$ |
| Xylenes+Toluene+20 ppm PTBS / Surfactant in salt water | $< 0.01 \pm 0.001$ |
| Xylenes+Toluene/20 ppm PAM in salt water | $< 0.01 \pm 0.001$ |
| Xylenes+Toluene/20 ppm PAM in Surfactant in salt water | $< 0.01 \pm 0.001$ |

**Migration of water into the oil phase studied using nectar drop** We examined the kinetics of the growth and the decay of the size of a nectar drop by placing a small (~ 0.6 μL) droplet on the tip of a glass bead. This bead was formed by melting a fine glass capillary tube and silanizing it with perfluorooctyl trichlorosilane. The small droplet of the sugar solution was deposited in the glass bead by bringing it in contact with a thin film (15 μm thickness) of sugar solution coated on a glass slide, followed by quick removal. This glass bead with the syrup drop was then dried in hot air and placed inside a plexi-glass chamber saturated with water vapor, the growth of which was observed under a microscope. The concentrated sugar solution was prepared by replicating as closely as possible the chemical composition of Agave nectar by adding 54 % (w/w) of D-Fructose (Fisher Science Education) and 18 % (w/w) of D-Glucose (Fisher Chemicals) in 28 % (w/w) of deionized (DI, Thermo Scientific Barnstead E-pure unit) water. The viscosity of this solution at room temperature (21 °C), as measured using an Ostwald viscometer, was found to be 13 mPa-s. The mixture was heated to 65 °C with constant stirring until a homogeneous solution was obtained. This drop hanging from the support was placed facing vertically down at a distance of ~1.2 mm from the interfacial region in the oil phase. At constant temperature and pressure, we can observe that the droplet size of the sugar drop grew larger in size based on the system of consideration.

**Molecular dynamics simulation.** Molecular dynamics simulations were performed using Large Atomic Molecular Massively Parallel Simulator (LAMMPS, ion source version 16Mar2018). Since this is an open access software, we do not see the need to repeat the related details that are already available publicly[58-63]. Nevertheless, we found it necessary to provide a few details that are relevant to this work, where dodecane was used as the organic phase, with dodecanoic acid (100 ppm), Brij-35 (100 ppm) and their mixture (50 ppm:50 ppm) as surfactants, and 0.6 M sodium chloride in water as the aqueous phase. The Optimized Potentials for Liquid Simulations,

All Atom (OPLS-AA) forcefield parameters were used. As for the non-bonded and bonded potential parameters of the various organic components, OPLS-AA forcefields were used. For DDA (dodecanoic acid) and Brij 35, the force field parameters were adapted from the publication of Shirley W. I. Siu et al[56]. The force fields and the partial charges of surfactant molecules as were obtained via quantum mechanical calculations were adapted from literature[53]. Finally, for water, a SPC/F$_w$ model, which is a 3 cite simple point-charge water model with flexible O-H bonds and H-O-H angle was adapted from literature[58]. Here, the energy of the initial structure of each component, as mentioned above, was first minimized using the steepest descent algorithm, followed by 1 ns pre-equilibration under the condition of constant pressure and temperature (NPT). Subsequent simulations were performed at a temperature of 298 K that was coupled to a Nose–Hoover algorithm using a 3d periodic boundary conditions (PBC). The short-range Coulomb interactions were calculated up to a distance of 1.2 nm with the long-range interactions calculated using the particle mesh Ewald (PME) reciprocal sum method. The short-range van der Waals (vdW) interactions were computed using the Lennard-Jones 12−6 potential that were calculated up to a distance of 1.2 nm, which accounts for the long-range dispersion correction for energy and pressure. All chemical covalent bonds were constrained by means of the Linear constraint solver (LINCS) algorithm.

The simulations (DPD-MD) combined dissipative particle dynamics (DPD) and molecular dynamics (MD), which were carried out for a duration of 300 ns using a simulation box with the lateral dimensions fixed at 10 nm x 10 nm laterally, while the vertical dimension changed according the stresses developed in the system, in accordance with the interfacial tensions $\gamma$. The interfacial tensions $\gamma = \frac{1}{2} L_z \left[ \langle p_{zz} \rangle - \frac{1}{2} \left( \langle p_{xx} \rangle + \langle p_{yy} \rangle \right) \right]$ was calculated using MD-DPD M106 source code (Version 2017.V1.2.3), where $L_z$ is the characteristic length of the interface in the $z$-axis and $\langle p_{xx} \rangle$, $\langle p_{yy} \rangle$, and $\langle p_{zz} \rangle$ are the time averaged diagonal components of the pressure tensor. These calculations yielded the surface tension of water as 72 mN/m, and the interfacial tensions of water against Dodecane and Decane as 52 mN/m, and 50.4 mN/m, respectively. The interfacial tensions of various interfaces are summarized in Table S2.

Returning to the details in performing the simulations, let us first emphasize that the radial distribution function $g(r)$ was calculated for both the equilibrium and nonequilibrium systems using the program RDF28 for each atom. This method characterizes the hydration behavior of each species and identifies the total number of H-bonds and the hydration number of any component[60,63]. In the radial distribution profiles, the characteristic peaks corresponding to the HB distance signifies whether the hydrogen bond forms between molecules of interest, namely dodecanoic acids, water and poly(oxyethylene) ether. The coordination numbers (CN) were estimated for each species by integrating the radial density function, up to a value that corresponds to its first minimum.

**Aggregation of Surfactant in various Solvents.** We extended the standard DPD simulations for estimating the repulsion parameter for different solvent systems in order to identify the interaction parameter for the surfactants in both organic and aqueous phases with oil and water being the respective implicit solvents. Figure 3 (D-F) shows the results of such a simulation where the self-assembling behaviors of DDA, Brij-35 and the mixture were studied with dodecane as implicit solvent. Similar studies with water as the implicit solvent are presented in Figure 7.

Let us now explain how we carried out the DPD-MD simulations. To begin with, a coarse grained DPD simulation was carried out for the surfactant molecules in the organic and water phases individually[59-61]. In order to carry out these simulations, all the particles were placed at random positions within the rectangular box of 100x100x100 nm$^3$ and using a grant size as 1 nm$^3$. These simulations allowed us to estimate the repulsion parameters (Table 4) between components. These new repulsion parameters were integrated into the MD simulation using LAMMPS, where we could obtain the atomistic radial distribution of the individual molecule. Figure 8C illustrates how these new parameters integrated with Non-equilibrium DPD-MD simulation (as discussed below) allowed the estimation of cross-sectional anatomy of various structures: vesicles, inverted emulsion and emulsion, which were performed in the NVE ensemble within a periodic boundary condition.

**Table S3.** Bead Notation and Repulsion Strength Parameters Used in the MD-DPD Simulation

| $a_{ij} = \frac{k_B T}{r_c}$ (eV/Å) | | | | | | | | | | |
|---|---|---|---|---|---|---|---|---|---|---|
| molecular prototype | bead name | P1 | P2 | P3 | D1 | D2 | B1 | B2 | E1 | W1 |
| PTBS (aliphatic tail) | P1 | 30 | 45 | 27 | 40 | 66 | 149 | 122 | 104 | 93 |
| PTBS (aromatic) | P2 | 68 | 52 | 70 | 24 | 108 | 74 | 25 | 35 | 26 |
| PTBS (aliphatic head group) | P3 | 27 | 70 | 25 | 51 | 70 | 65 | 115 | 32 | 60 |
| Carboxylic acid group | D1 | 35 | 24 | 39 | 54 | 89 | 84 | 47 | 45 | 145 |
| Aliphatic tail (DDA) | D2 | 80 | 107 | 50 | 58 | 15 | 40 | 167 | 154 | 125 |
| Ethylene oxide group (Brij-35) | B1 | 187 | 45 | 104 | 72 | 48 | 63 | 45 | 70 | 57 |
| Aliphatic tail (Brij-35) | B2 | 145 | 38 | 115 | 47 | 125 | 82 | 54 | 43 | 45 |
| Dodecane | E1 | 109 | 158 | 89 | 21 | 154 | 65 | 12 | 25 | 30 |
| Water | W1 | 50 | 15 | 89 | 78 | 135 | 43 | 45 | 30 | 25 |

Having discussed about the specifics of the DPD-MD simulation above, we now move to the DPD-MD simulations for the non-equilibrium emulsion formation kinetic studies. These non-equilibrium simulations were performed with periodic boundary conditions under a non-isothermal ensemble. The reaction field corrections were applied for compensating the mean effect of electrostatic interactions beyond 1.2 nm distance, using a dielectric permittivity of 60 as appropriate for the SPC water model. All the systems were ensured to have an effective solvent density by controlling the volume ratios of the solute to the solvent. The simulations were carried out for a duration of 300 ns (as mentioned above) with the trajectory capture of 20 ns for further analysis.

In the Non-equilibrium DPD-MD simulation, we obtained the trajectory of a system as a function of time in an NVE ensemble. The system traverses through multiple microstates before it finally reaches an equilibrium. While an equilibrium DPD-MD simulation was performed with a widely used Nose-Hover thermostat, this does not have the capability to ascertain the path of the system that reaches the final state when the system is coupled with the hydrodynamic forces. We used DPD-MD simulations with a non-isothermal algorithm to understand the mechanism of emulsification. This NVE ensemble fails to converge within the scope of the first-fluctuation dissipation theorem. Hence these simulations were performed with a modified DPD algorithm (using Second fluctuation dissipation theorem allowing the kinetic temperature to evolve). A velocity verlet method is also adopted to these simulations in order to prevent the self-penetration of the molecules within each other.

**Ionization strength – information from MD simulations.** Poisson-Boltzmann (PB) method was used in order to ensure that the protonation and de-protonation strengths have been estimated correctly in the above NEMD simulations. This method combined with DPD-MD simulations were used for analyzing the anatomy of the components of emulsification. In the estimation of ionization strength, the artifacts due to multiple bond formation in a head - head interaction were eliminated by constraining the bond angle in the system by standard method using LINCS algorithm[62-63]. In the aggregation of surfactants in various solvents we found that the DDA forms lateral bonds with the neighboring DDA molecules. It also forms head on bonding with Brij-35 molecule and water molecule.

**Table S4**. Interfacial tension and Molecular Surface area estimate using MD-DPD simulation with a Grain size of 0.1 nm$^3$.

| Component | Interfcial Tension, mN/m | Total Area A (nm²) | $N_{DDA}$ | $N_{Brij}$ | $\frac{N_{DDA}}{N_{Brij}}$ | $\frac{A}{N_{DDA}+N_{Brij}}$ nm² |
|---|---|---|---|---|---|---|
| Flat interface | 7.6 | 10,000 | 6745 | 6012 | 1.12 | 0.78 |
| Oil in Water Emulsion | 1.4 | 493 | 950 | 650 | 1.46 | 0.31 |
| Water in Oil Emulsion | 2.1 | 1319 | 1901 | 1450 | 1.31 | 0.39 |
| Vesicle | 2.3 | 1256 | 1865 | 1542 | 1.21 | 0.37 |

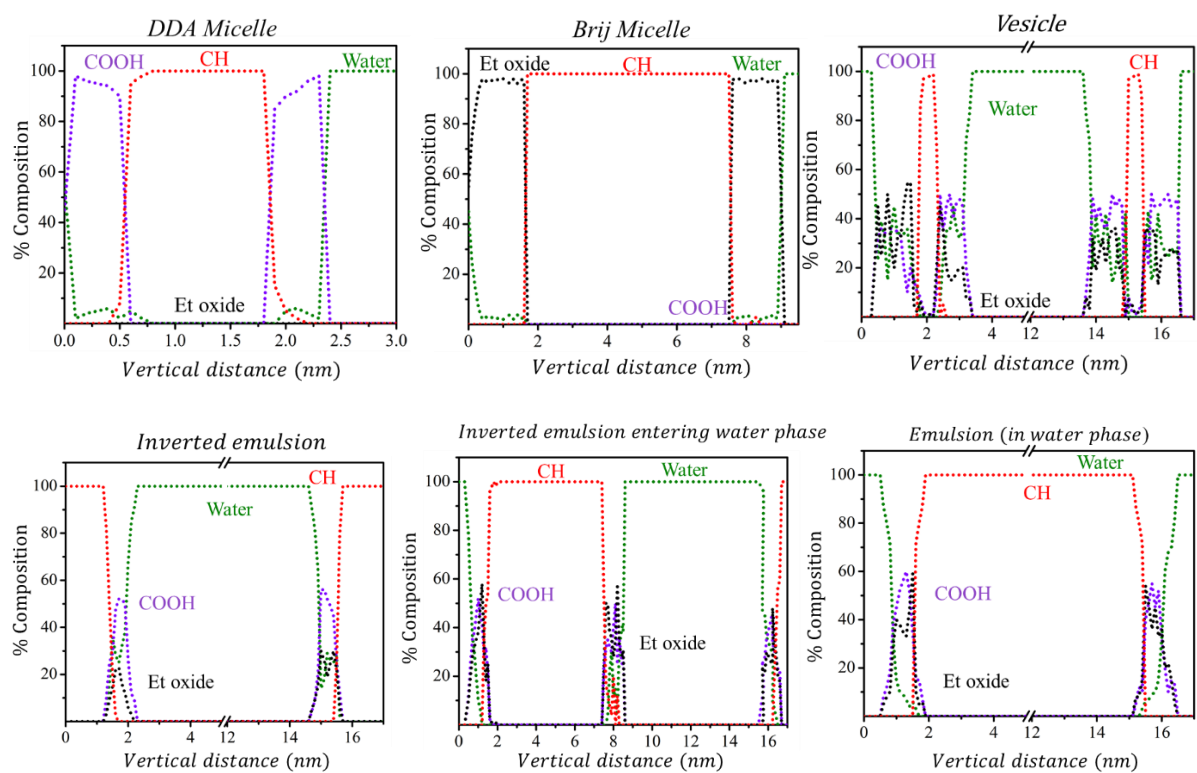

**Figure S10.** The anatomy of the vesicle like structures can be visualized when the concentrations of water, hydrocarbon, carboxylic acid and water are plotted across it. Along with these structures,

the composition profile of these groups is shown for a micelle, vesicle, water-in-oil and oil-water emulsions.

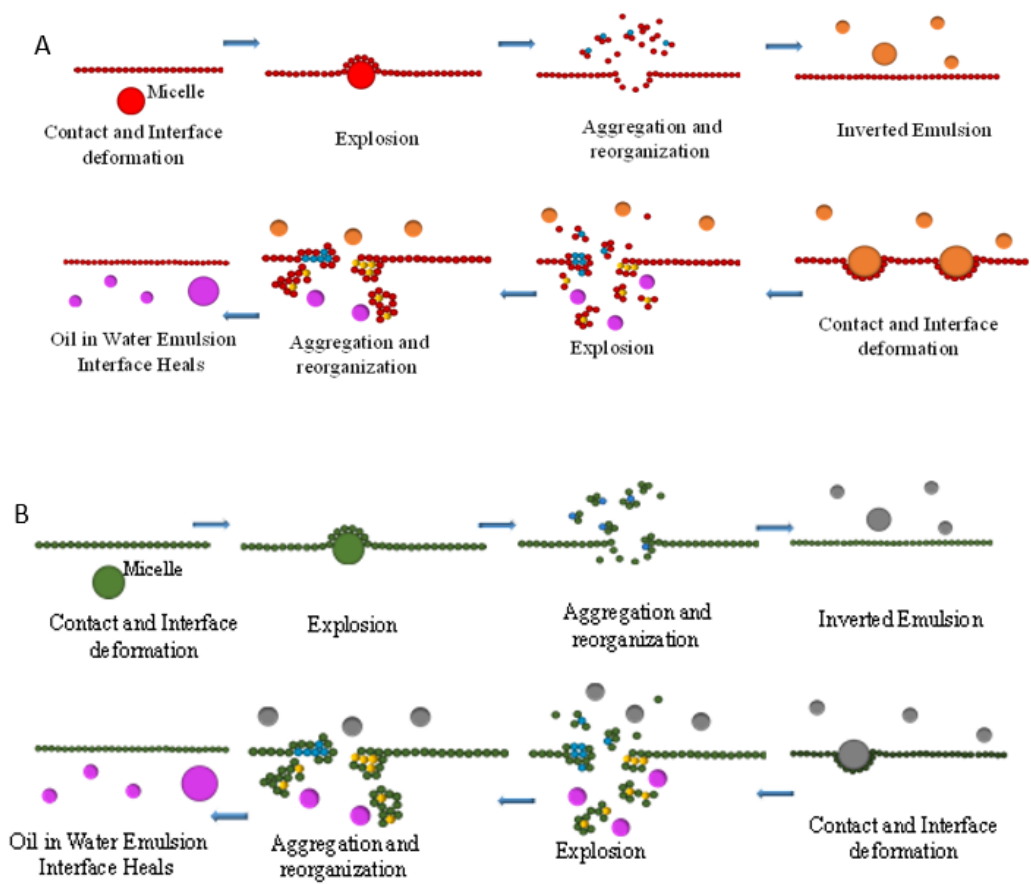

**Figure S11. (A and B)** Formation of water-in oil and oil-in-water emulsions. When the (A) dodecanoic acid (A) and Brij-35 (B) form micelles in the aqueous phase, it eventually comes into contact with the interface and explodes. The fragments resulting from this explosion cluster together and forms the inverted (i.e. water-in-oil) emulsion. When such an inverted emulsion comes into contact with the interface, another explosion occurs, following which the fragments cluster together and transform into oil-in-water emulsions. (Note: the color coding for different components are as follows, Red: DDA, Green: Brij-35, Blue: water, Yellow: Oil, Grey and Orange: inverted emulsion and Pink: Emulsion)

**Table S5a**. Diffusivity of emulsion as obtained from Figure 10 is plotted against the value $(D_e)$ obtained from the Stokes-Einstein relation: $D_e = k_B T/(6\pi\eta r)$ , where $k_B$ is the Boltzmann constant, $T$ is the temperature, $\eta$ is the viscosity of the aqueous phase. $r$ is the average radius of the droplet, which was estimated using CHDF (capillary hydrodynamkc flow chromatography, Figure S10.

| Organic phase | $r$ $(m)$ | $D$, $m^2/s$ | $D_e$, $m^2/s$ | $D/De$ |
|---|---|---|---|---|
| Dodecane | $1.10(\pm0.002)\times10^{-7}$ | $2.2(\pm0.87)\times10^{-10}$ | $1.81\times10^{-12}$ | $1.22\times10^2$ |
| Decane | $1.00(\pm0.05)\times10^{-7}$ | $6.2(\pm0.25)\times10^{-10}$ | $1.99\times10^{-12}$ | $3.12\times10^2$ |
| Octane | $2.50(\pm0.12)\times10^{-8}$ | $4.6(\pm0.12)\times10^{-10}$ | $7.95\times10^{-12}$ | $5.79\times10^1$ |
| Toluene | $1.30(\pm0.09)\times10^{-7}$ | $1.6(\pm0.01)\times10^{-10}$ | $1.53\times10^{-12}$ | $1.05\times10^2$ |
| Styrene | $1.50(\pm0.24)\times10^{-7}$ | $3.8(\pm0.02)\times10^{-10}$ | $1.32\times10^{-12}$ | $2.87\times10^2$ |
| Xylenes | $5.00(\pm0.14)\times10^{-8}$ | $1.0(\pm0.08)\times10^{-11}$ | $3.97\times10^{-12}$ | $2.52\times10^0$ |

**Table S5b.** Diffusivity of emulsion as obtained from figure 10 is plotted against the value ($D_e$) obtained from the Stokes-Einstein relation: $D_e = k_B T/(6\pi\eta r)$ , where $k_B$ is the Boltzmann constant, $T$ is the temperature, $\eta$ is the viscosity of the aqueous phase. $r$ is the average radius of the droplet, which was estimated using CHDF (capillary hydrodynamkc flow chromatography, figure S9. The ratio

| System | $r$ $(m)$ | $D$, $m^2/s$ | $D_e$, $m^2/s$ | $D/De$ |
|---|---|---|---|---|
| Diesel w/ 500 ppm of S (●) | $1.17(\pm0.102)\times10^{-7}$ | $1.67(\pm0.25)\times10^{-4}$ | $1.70\times10^{-12}$ | $9.84\times10^7$ |
| Diesel w/ 300 ppm of S (●) | $4.14(\pm0.215)\times10^{-8}$ | $1.30(\pm0.001)\times10^{-4}$ | $4.80\times10^{-12}$ | $2.71\times10^7$ |
| Diesel w/ 200 ppm of S (●) | $2.67(\pm0.012)\times10^{-7}$ | $1.10(\pm0.26)\times10^{-4}$ | $7.45\times10^{-13}$ | $1.48\times10^8$ |
| Diesel w/ 100 ppm of S (●) | $1.11(\pm0.202)\times10^{-7}$ | $9.15(\pm0.36)\times10^{-5}$ | $1.79\times10^{-12}$ | $5.13\times10^7$ |
| Diesel w/ 50 ppm of S (●) | $5.40(\pm0.004)\times10^{-7}$ | $1.90(\pm0.15)\times10^{-5}$ | $3.68\times10^{-13}$ | $5.17\times10^7$ |
| Diesel + 20 ppm PTBS w/ 100ppm of S (◯) | $3.26(\pm0.001)\times10^{-8}$ | $5.00(\pm0.12)\times10^{-9}$ | $6.10\times10^{-12}$ | $8.20\times10^2$ |
| Diesel + 40 ppm PTBS w/ 100 ppm of S (◯) | $1.16(\pm0.025)\times10^{-7}$ | $1.40(\pm0.01)\times10^{-9}$ | $1.71\times10^{-12}$ | $8.20\times10^2$ |
| Diesel + 20 ppm PTBS w/ 500ppm of S (◯) | $1.00(\pm0.010)\times10^{-8}$ | $1.63(\pm0.01)\times10^{-4}$ | $1.99\times10^{-11}$ | $8.20\times10^6$ |
| Diesel + 40 ppm PTBS w/ 500ppm of S (◯) | $2.42(\pm0.122)\times10^{-9}$ | $1.55(\pm0.01)\times10^{-4}$ | $8.22\times10^{-11}$ | $1.89\times10^6$ |
| Diesel + 80 ppm PTBS w/ 500ppm of S (◯) | $1.16(\pm0.090)\times10^{-7}$ | $1.50(\pm0.01)\times10^{-4}$ | $1.71\times10^{-11}$ | $8.78\times10^6$ |
| Diesel + 100 ppm PTBS w/ 500ppm of S (◯) | $8.00(\pm0.325)\times10^{-9}$ | $1.45(\pm0.01)\times10^{-4}$ | $2.48\times10^{-11}$ | $5.84\times10^6$ |
| Benzene w/ 500 ppm of S (◻) | $1.18(\pm0.122)\times10^{-8}$ | $1.68(\pm0.01)\times10^{-4}$ | $1.68\times10^{-11}$ | $1.00\times10^7$ |
| Toluene w/ 500 ppm of S (◻) | $3.49(\pm0.101)\times10^{-7}$ | $8.75(\pm0.01)\times10^{-5}$ | $5.69\times10^{-13}$ | $1.24\times10^8$ |
| Styrene w/ 500 ppm of S (◻) | $1.35(\pm0.08)\times10^{-7}$ | $7.50(\pm0.16)\times10^{-5}$ | $1.48\times10^{-12}$ | $5.08\times10^7$ |
| Xylenes w/ 500 ppm of S (◻) | $9.13(\pm0.133)\times10^{-8}$ | $5.00(\pm0.25)\times10^{-5}$ | $2.18\times10^{-12}$ | $2.30\times10^7$ |
| Octane w/ 500 ppm of S (◇) | $1.11(\pm0.24)\times10^{-8}$ | $8.50(\pm0.46)\times10^{-5}$ | $1.79\times10^{-11}$ | $4.74\times10^6$ |
| Decane w/ 500ppm of S (◇) | $9.29(\pm0.43)\times10^{-8}$ | $4.00(\pm0.45)\times10^{-5}$ | $2.14\times10^{-12}$ | $1.87\times10^7$ |
| Dodecane w/ 500ppm of S (◇) | $3.00(\pm0.05)\times10^{-7}$ | $2.00(\pm0.37)\times10^{-5}$ | $6.62\times10^{-13}$ | $3.02\times10^7$ |
| Benzene w/ 100 ppm of S (▲) | $2.56(\pm0.53)\times10^{-8}$ | $1.50(\pm0.11)\times10^{-4}$ | $7.75\times10^{-12}$ | $1.94\times10^7$ |
| Cyclohexane w/ 100 ppm of S (◇) | $2.80(\pm0.126)\times10^{-7}$ | $2.20(\pm0.87)\times10^{-6}$ | $7.10\times10^{-13}$ | $3.10\times10^6$ |

**Table 5b (continued)**

| System | r (m) | D, m²/s | D_e, m²/s | D/De |
|---|---|---|---|---|
| Cyclohexane+Dodecane w/ 100 ppm of S (◇) | $1.50(\pm0.002)\times10^{-7}$ | $2.00(\pm0.13)\times10^{-6}$ | $1.32\times10^{-12}$ | $1.51\times10^{6}$ |
| Diesel w/ 20 ppm of PAM with 100 ppm of S (▲) | $2.62(\pm0.04)\times10^{-8}$ | $2.00(\pm0.11)\times10^{-8}$ | $7.57\times10^{-12}$ | $2.64\times10^{3}$ |
| Diesel w/ 40 ppm of PAM with 100 ppm of S (▲) | $8.80(\pm0.46)\times10^{-8}$ | $2.00(\pm0.10)\times10^{-9}$ | $2.26\times10^{-12}$ | $8.86\times10^{2}$ |
| Benzene+Toluene w/ 100 ppm of S (■) | $1.20(\pm0.16)\times10^{-7}$ | $3.50(\pm0.11)\times10^{-6}$ | $1.66\times10^{-12}$ | $2.11\times10^{6}$ |
| Benzene+Styrene w/ 100 ppm of S (□) | $2.90(\pm1.34)\times10^{-7}$ | $2.50(\pm0.25)\times10^{-6}$ | $6.85\times10^{-13}$ | $3.65\times10^{6}$ |
| Benzene+Xylenes w/ 100 ppm of S (□) | $1.10(\pm0.21)\times10^{-7}$ | $1.20(\pm0.59)\times10^{-6}$ | $1.81\times10^{-12}$ | $6.64\times10^{5}$ |
| Simulated mixture w/ 100 ppm of S (○) | $2.25(\pm0.536)\times10^{-7}$ | $2.20(\pm0.46)\times10^{-6}$ | $8.83\times10^{-13}$ | $2.49\times10^{6}$ |
| Equal volume mixture of all solvents w/ 100 ppm of S (◎) | $2.70(\pm0.26)\times10^{-7}$ | $5.60(\pm0.13)\times10^{-6}$ | $7.36\times10^{-13}$ | $7.61\times10^{6}$ |
| Benzene+Ethanol w/ 100 ppm of S (☆) | $4.00(\pm0.33)\times10^{-7}$ | $4.00(\pm0.06)\times10^{-5}$ | $4.97\times10^{-13}$ | $8.05\times10^{7}$ |
| Toluene+Ethanol w/ 100 ppm of S (☆) | $6.54(\pm0.01)\times10^{-9}$ | $8.00(\pm0.01)\times10^{-4}$ | $3.04\times10^{-11}$ | $2.63\times10^{7}$ |
| Styrene+Ethanol w/ 100 ppm of S (☆) | $1.00(\pm0.08)\times10^{-8}$ | $5.00(\pm0.13)\times10^{-4}$ | $1.99\times10^{-11}$ | $2.52\times10^{7}$ |
| Xylenes+Ethanol w/ 100 ppm of S (☆) | $3.03(\pm0.015)\times10^{-9}$ | $4.65(\pm0.09)\times10^{-4}$ | $6.55\times10^{-11}$ | $7.09\times10^{6}$ |
| Octane+Ethanol w/ 100 ppm of S (☆) | $3.80(\pm0.01)\times10^{-9}$ | $3.00(\pm0.37)\times10^{-4}$ | $5.23\times10^{-11}$ | $5.74\times10^{6}$ |
| Decane+Ethanol w/ 100 ppm of S (☆) | $9.56(\pm0.01)\times10^{-9}$ | $4.52(\pm0.23)\times10^{-4}$ | $2.08\times10^{-11}$ | $2.17\times10^{7}$ |
| Dodecane+Ethanol w/ 100 ppm of S (☆) | $1.16(\pm0.002)\times10^{-8}$ | $5.16(\pm0.01)\times10^{-4}$ | $1.71\times10^{-11}$ | $3.02\times10^{7}$ |
| Cyclohexane+Dodecane w/ 20 ppm of PAM with 100 ppm of S (▲) | $4.89(\pm0.02)\times10^{-8}$ | $1.50(\pm0.08)\times10^{-8}$ | $4.06\times10^{-12}$ | $3.69\times10^{3}$ |
| Cyclohexane+Dodecane /40 ppm of PAM with 100 ppm of S (▲) | $2.80(\pm0.54)\times10^{-7}$ | $1.50(\pm0.01)\times10^{-9}$ | $7.10\times10^{-13}$ | $2.11\times10^{3}$ |
| Cyclohexane+Dodecane+20 ppm PTBS w/ 100 ppm of S (▲) | $4.00(\pm0.71)\times10^{-7}$ | $2.50(\pm0.68)\times10^{-6}$ | $4.97\times10^{-13}$ | $5.03\times10^{4}$ |
| Cyclohexane+Dodecane+40 PPM PTBS w/100 ppm of S (▲) | $3.90(\pm0.437)\times10^{-7}$ | $5.00(\pm0.13)\times10^{-9}$ | $5.10\times10^{-13}$ | $9.81\times10^{3}$ |
| Dodecane+Decane w/ 100 ppm of S (▲) | $1.40(\pm0.002)\times10^{-7}$ | $6.61(\pm0.87)\times10^{-5}$ | $1.42\times10^{-12}$ | $4.66\times10^{7}$ |
| Xylenes+Toluene w/ 100 ppm of S (▲) | $1.50(\pm0.047)\times10^{-7}$ | $1.21(\pm0.89)\times10^{-5}$ | $1.32\times10^{-12}$ | $9.13\times10^{6}$ |